\documentclass[twocolumn,showpacs,floatfix,pre,superscriptaddress,numbers,sort&compress]{revtex4-1}
\RequirePackage{lineno}

\usepackage{subfiles}
\usepackage{graphicx}
\usepackage{dcolumn}
\usepackage{epsf} 
\usepackage{amsmath}
\usepackage{mathrsfs}
\usepackage{euscript}
\usepackage{calrsfs}
\usepackage{bm} 
\usepackage{setspace}
\usepackage[dvipsnames]{xcolor}
\usepackage[toc,page]{appendix}
\usepackage{natbib}

\newcommand{\ck}{\color{black} }

\begin{document}


\title{
Collineations of particles in the Kob-Andersen system}

\author{V.A.~Levashov}
\affiliation{Technological Design Institute of Scientific Instrument Engineering, 630055, Novosibirsk, Russia. E-mail: valentin.a.levashov@gmail.com}


\begin{abstract}
Numerous indications suggest that subtle changes 
occurring in the structures of liquids on 
supercooling are connected to the phenomenon 
of the glass transition and that 
detailed understanding of these changes 
is crucial for the development 
of new glasses with desired properties.
J.D. Bernal in his 1962 Bakerian lecture, 
in particular, 
reported about an observation of approximately 
linear chains of several particles, 
referred to as collineations.
He found that in the studied hard sphere system, 
these collineations can contain up to eight particles. 
Since then, 
the collineations of three particles have been discussed 
in many papers 
in the context of the splitting of the second peak 
in pair density functions of supercooled liquids and glasses.
However, it appears that longer collineations involving 
more that three particles have not been systematically studied.  
Here, we report on our study of such collineations for the Kob-Andersen system of particles on cooling for the parent and inherent structures. Contrary to intuition, our findings reveal that below the potential energy landscape crossover temperature, the number of collineations in the parent structures can exceed that of the corresponding inherent structures.
We also introduce a model that connects long collineations with the pair density and angular density distribution functions and demonstrate that this model describes long collineations quite well.
The second part of the paper explores potential connections between collineations and: 1) the disclination lines associated with the geometric frustration approach, 2) low-energy clusters from the topological cluster classification approach, 3) chain-like cooperative motion of particles in low-temperature supercooled liquids.
For the studied system, according to the used methods, no clear connection was found between collineations and these phenomena.\\
\end{abstract}



\today
\maketitle


\section{Introduction}\label{s:intro2}

The investigations of the structural organization of 
supercooled liquids and glasses continue to garner
significant attention  
despite decades of studies 
\cite{tanaka2022roles,tanaka2019revealing,
kelton2023perspective,greer2009metallic,
royall2015role,cheng2011atomic,kauzmann1948nature}. 
Understanding of structures of 
such non-crystalline materials
is important for unraveling 
the nature of the glass transition (GT) 
and developing new glasses with desired 
properties \cite{kelton2023perspective,greer2009metallic}.
It is generally recognized that structural changes 
occurring in liquids on supercooling and approach 
of the GT are subtle while changes 
in the dynamical properties are tremendous 
\cite{tanaka2022roles,tanaka2019revealing,kelton2023perspective,
greer2009metallic,royall2015role,cheng2011atomic,kauzmann1948nature}. 
Nevertheless, some relatively minor structural changes clearly correlate with the GT. 
For example, 
from a macroscopic perspective, 
there is a change in the slope 
at the GT in the temperature dependence 
of the liquid's volume at constant 
pressure \cite{kauzmann1948nature}. 
Additionally, 
from an average microscopic standpoint, 
there is a change in the slope of the ratio of the depth 
of the first minimum to the height of the first maximum 
in the pair density function (PDF) \cite{wendt1978empirical}. 
Another microscopic feature 
often considered as a signature of the glass state 
is the splitting of the second peak 
in the PDF \cite{royall2015role,cheng2011atomic,
finney1970random, bennett1972serially, 
wendt1978empirical, voloshin1997origin, 
truskett1998structural,o2005structure,
luo2006pair,pan2011origin,ding2015second,wu2015hidden}. 
The development of this feature indicates that 
ordering on medium-range length scales is relevant 
to the GT \cite{voloshin1997origin,truskett1998structural,o2005structure,luo2006pair,pan2011origin,ding2015second}. 
The structural organization associated 
with the splitting of the second peak is 
directly related to the phenomenon studied 
in this paper, as we discuss later.
Observations of the aforementioned features in different systems 
have led to extensive efforts to interpret them in terms of structural changes. 
Of particular importance is the observation that, 
in many liquids upon supercooling, icosahedral ordering develops. 
It is often assumed that icosahedral ordering prevents crystallization 
due to its five-fold symmetry, which is incompatible with crystalline periodic order ~\cite{royall2015role,frank1958complex,kleman1979tentative,sadoc1982order, nelson1983order,steinhardt1983bond, jonsson1988icosahedral,derlet2020correlated,qi1991icosahedral, tarjus2005frustration,pedersen2010geometry,derlet2020correlated}. 
Furthermore, it has been demonstrated that icosahedra in such liquids 
have a tendency to agglomerate into domains \cite{dzugutov2002decoupling,coslovich2007locally,malins2013lifetimes}. 
This indicates that in liquids upon supercooling, 
structural heterogeneities develop with characteristic 
length scales larger than interparticle distances.
The development of icosahedral ordering 
on the length scales associated with a central particle and its nearest neighbors does not occur in all liquids. 
This is because, in some liquids, the lowest energy local structure is not an icosahedron \cite{tanaka2019revealing,royall2015role,malins2013lifetimes}. 
However, 
recent studies have demonstrated that in such liquids, 
five-fold structural symmetry may still be present 
on the length scales associated with the second 
and further neighbors \cite{fan2020unveiling,zhang2020revealing}. 
This once again underscores the importance 
of structural medium-range order for the GT.

While the importance of medium-range order 
has been discussed in many publications, 
the structural descriptors commonly employed 
to address the evolution of the structure are typically local. 
For instance, 
Voronoi polyhedra, 
bond orientational order parameters, 
and the numbers of common neighbors for neighbor pairs 
are frequently used ~\cite{royall2015role,gellatly1982characterisation, steinhardt1983bond,jonsson1988icosahedral,honeycutt1987molecular, stukowski2012structure,malins2013identification, malins2013lifetimes}. 
Subsequently, in addressing structural changes upon supercooling, 
the ways in which these units combine are often 
considered \cite{tanaka2019revealing,ding2015second, fan2020unveiling,royall2015role, malins2013lifetimes,coslovich2007understanding}.

Due to the absence of periodicity in liquids, 
structural changes occurring during supercooling 
cannot be observed using conventional crystallographic techniques \cite{tanaka2022roles,tanaka2019revealing,kelton2023perspective,greer2009metallic,royall2015role, cheng2011atomic}. 
Consequently, the PDF, 
which is experimentally accessible for liquids, 
often serves as the standard data for structural analysis. 
However, the PDF measures the angular average of the structural information, 
and this information is also averaged over a length scale significantly 
larger than the interatomic distance. 
On the other hand, simulations of liquid structures and studies of other systems, 
such as hard spheres, clearly indicate that the structures of liquids 
are inhomogeneous at the atomic scale. 
For these reasons, accurate interpretations 
of the PDF require careful modeling.
It is worth noting in this context that 
the average PDF for some glasses can 
be modeled with relatively complex 
crystal structures \cite{gaskell1978new,miracle2004structural}, 
notwithstanding the structural 
heterogeneities mentioned earlier.

As we already noted previously, 
a particular feature of the PDF that has attracted 
significant attention is the splitting of the second peak. 
This phenomenon has been observed in various systems, 
including hard-sphere, 
soft-sphere, 
Lennard-Jones systems, metallic glasses,
and colloidal 
systems \cite{finney1970random,simpson1972bubble,raveche1976computer,
polk1973dense,yamamoto1978structural,liu2010metallic,wu2015hidden}. 
The prominence of the second peak splitting increases below the GT, 
and historically, this feature has been utilized as a structural 
indicator of the glass state \cite{wendt1978empirical,cheng2011atomic}. 
Additionally, it has been observed that 
the relative heights of the two peaks vary 
depending on the specific system and the method of 
preparation \cite{o2005structure,cheng2011atomic,pan2011origin,ding2015second,wu2015hidden}.

In the early studies, 
it was observed that the two subpeaks of 
the second peak in the PDF occur 
at distances of approximately $\sim 1.7a$ and $\sim 2a$, 
where $a$ represents the position of the first peak in the PDF, i.e., 
the nearest neighbor distance. 
The distance $1.7a$ is roughly equivalent 
to twice the height of an equilateral triangle with side $a$. 
Consequently, this distance has been associated with 
the in-plane arrangement of four particles into 
two equilateral triangles with a common edge 
(distance $\sqrt{3}a \approx 1.73 a$). 
Another potential geometry corresponding to the first part of 
the split second peak involves two regular tetrahedra with 
a common face (distance $2\sqrt{2/3}a \approx 1.63 a$). 
The position of the second part of the split second peak occurs at 
a distance of $\sim 2a$, 
which is clearly associated with 
the collineations of triplets of particles \cite{finney1970random}.
In later studies it was 
suggested that the splitting of the second peak is a signature 
of formation tetrahedral or crystal-like domains Ref.~\cite{voloshin1997origin,truskett1998structural,o2005structure}, 
Recent investigations into 
the structural origin of the splitting of the second peak in the PDF 
have typically focused on studying the types of nearest neighbor 
clusters for the second neighbors, 
with particular emphasis on the number of their common 
neighbors \cite{pan2011origin,ding2015second,luo2006pair}. 
For the case of the Kob-Andersen (KA) system studied in this paper, 
such investigation had been conducted in Ref.~\cite{luo2006pair}. 
An alternative approach to addressing the splitting 
of the second PDF peak, based on the consideration 
of rings of particles, 
has been explored in Ref.~\cite{o2005structure}.
In 1962, J.D. Bernal in his Bakerian lecture reported 
about observation of collineations involving more 
than three particles \cite{bernal1964bakerian}. 
Clearly, such collineations are associated 
with the splitting of the second peak 
in the PDF are related 
to the nature of intermediate range order. 
Bernal, in particular, noticed collineations 
involving up to 8 particles, while the average length of 
such collineations was estimated as 4. 
Bernal also noted that the existence of collineations 
seems to align with the shape of 
the angular distribution function (ADF) presented 
in Ref.~\cite{scott1964angular}. 
Furthermore, Bernal suggested that collineations 
might play a role in the diffusion process. 
It was proposed that if there is a sufficiently 
large hole at one end of a collineation, 
then the entire collineation can move collectively. 
Thus, a hole at one end of 
the collineation would reappear 
at another end. 
This suggestion bears resemblance to later 
observations of string-like cooperative motion \cite{donati1998stringlike}. 
However, as we demonstrate below, these two phenomena do not appear to be related.
Since 1962, Bernal’s findings regarding collineations
of multiple particles have been mentioned in more 
than just a few publications, 
including Ref.~\cite{finney1970random,simpson1972bubble,polk1973dense, raveche1976computer,yamamoto1978structural,
popescu1984defect,rafizadeh1990mechanical,
kleman1989curved,shcherbak1999pre, finney2013bernal}. 
For instance, in Ref.~\cite{popescu1984defect}, 
there is a suggestion that collineations might 
be connected to the disclination lines in the Frank-Kasper 
and geometrical frustration approaches. 
It is also reasonable to mention Ref.~\cite{liu2010metallic} 
in the context of collineations. 
It has been suggested in Ref.~\cite{liu2010metallic} that structural
organization of metallic glasses on the intermediate range can be considered
as a combination of spherical periodic order and local translational symmetry.
Further, it was argued that local translational symmetry may be  one-dimensional.
It is possible that collineations of particles might represent this
one-dimensional local translational symmetry.
However, 
to the best of our search efforts, 
long collineations do not appear to have been systematically studied.
In our investigation of the binary $(80\%:20\%)$ 
KA system of particles, we observed the formation 
of long collineations formed by the larger particles ($80\%$) 
and decided to explore this phenomenon.

This paper effectively is divided into two parts. 
In the first part, 
we present our statistical investigations of collineations 
on both the parent and inherent 
structures \cite{1998SastryPEL,debenedetti2001supercooled}
and describe how to relate the PDF and
the ADF to the collineations
and demonstrate that the suggested relation works quite well.
In the second part, we describe our investigations 
of possible connections of the collineations with some other phenomena.
In particular, our study investigated whether collineations are associated with:
1)the major skeleton of the Frank-Kasper structure 
or the disclination lines associated with the geometric frustration approach,
2) some clusters identified by the topological cluster classification approach,
3) strings composed of the most mobile particles.
Across all considered cases, our findings, based on the applied methods, indicate that particles forming collineations are not associated with any of the considered groups.
In the aforementioned considerations, 
we compared the average diffusion rate of particles 
forming the collineations with those of all large particles. Our findings show that, on average, 
particles forming the collineations exhibit 
a slightly slower diffusion rate 
compared to the average large particle.
In recent years, 
the role of developing intermediate range order 
in dynamic slowdown and the GT 
has captured significant attention \cite{malins2013lifetimes,wu2015hidden,sheng2006atomic, pan2011origin,berthier2012static,ryu2021medium,ding2015second, tah2022kinetic,levashov2017contribution,fan2020unveiling, zhang2020revealing,yuan2021connecting}
Collineations can also be considered as a manifestation of developing intermediate range order. 
It is evident that average short-range order should 
geometrically align with the average intermediate range 
order and beyond. 
Thus, it challenging to determine 
which structural changes are the cause 
and which are the consequence.
Yet, 
from our perspective, 
a specific feature observed in the behavior 
of the ADF suggests that the development 
of intermediate range order and beyond is 
the driver for certain changes 
in the short-range order.

The paper is structured as follows.

In Section \ref{sec:simulation-details}, 
we provide a brief overview of the simulation 
procedure and data analysis, 
with more detailed information available 
in the supplemental materials (SM).

Sections \ref{sec:chains-def} and \ref{ssec:observation-of-chains} 
define and describe the observed collineations.
In Sections \ref{sec:Analysis-of-PDF-and-ADF} and \ref{sec:PDF-ADF-Chains}, we delve into the behaviors of the PDF and ADF. 
We then propose a relation that connects these functions with the collineations, demonstrating its effectiveness in describing the data.
Then, in sections \ref{sec:Discussion-Frank-Kasper}, \ref{sec:Discussion-TCC}, and \ref{sec:Discussion-Chain-Like}, we explore potential connections of the collineations with 
the geometrical frustration approach, 
selected clusters from the topological cluster classification method, 
and the stringlike cooperative motion. 
We provide our concluding remarks in Section \ref{sec:conclusion}.

\section{The used potential and simulations details \label{sec:simulation-details}}

In this paper, we studied the KA model system 
has been extensively investigated previously 
\cite{1995KobAndersen01,1995KobAndersen02,kob1997dynamical,
donati1998stringlike,1998SastryPEL,2002KApressureBagchi,
biroli2008thermodynamic,berthier2009nonperturbative,malins2013lifetimes,
malins2013identification,coslovich2007locally,coslovich2013static,
zhang2020revealing,das2022crossover}.
In particular, 
there were studies of the short and medium range order
in this system
\cite{1995KobAndersen01,1995KobAndersen02,kob1997dynamical,
donati1998stringlike,1998SastryPEL,2002KApressureBagchi,
biroli2008thermodynamic,berthier2009nonperturbative,malins2013lifetimes,
malins2013identification,coslovich2007locally,coslovich2013static,
zhang2020revealing,boattini2021averaging}).
In our computer simulations, 
we used the modified forms 
\cite{stoddard1973numerical} 
of the KA potentials \cite{1995KobAndersen01,1995KobAndersen02}. 
The unmodified KA potentials are 
the shifted Lennard-Jones potentials 
with the parameters of length, 
$\sigma_{ab}$, and energy, $\epsilon_{ab}$, 
chosen in a particular way.
The modified potentials go 
to zero at the cutoff distances 
$r_{abc} = 2.5\sigma_{ab}$ 
with zero derivatives (forces). 
The same modified potentials have been used, 
for example, in 
Ref.\cite{stoddard1973numerical,1998SastryPEL,
2002KApressureBagchi,malins2013lifetimes}.
We provide more details 
on the used potentials 
and the simulation procedure 
in the supplemental materials (SM).
Masses of all particles 
are assumed to be the same, $m$.
The chosen units of 
length, energy and time are:
$\sigma \equiv \sigma_{AA}$, $\epsilon = \epsilon_{AA}$,
$\tau \equiv \sqrt{m \sigma^2/\epsilon}$.
We performed NVT simulations 
mostly on the system 
of 8000 particles 
(6400 $A$-particles and 1600 $B$-particles). 
The average density of the particles 
was $\rho_o = 1.2 \sigma^{-3}$.  
We also performed some of the simulations on the system containing $64\cdot 10^3$ particles. 
We used the LAMMPS molecular dynamics (MD) 
program \cite{Plimpton1995,thompson2022lammps,lammps}.
The used value of the time step, $dt$, 
varied from $0.0005\tau$ at $T=6.0\epsilon$
up to $0.005\tau$ at $T<0.5\epsilon$.
At every considered temperature 
we performed 
(10 for the systems of 8000 particles)/(6 for the system of $64 \cdot 10^3$ particles) simulations 
starting from high-$T=6.0\epsilon$ independent structures.
In each of these independent simulations, 
we accumulated 
(100 for the systems of 8000 particles)/(20 for the system of $64 \cdot 10^3$ particles) 
configurations.
The final configurations of particles 
from the previous higher temperature 
(after the relaxation)
were used as initial configurations for 
relaxation at the next lower temperature.
The initial relaxation time for all 
$T>0.46\epsilon$ was larger than $10\tau_{\alpha}(T)$
at each studied temperature. 
Here, $\tau_{\alpha}(T)$ is the relaxation time of the
intermediate self-scattering function at temperature $T$. 
Then, the configurations were saved 
consequently with time intervals that
were at least several $\tau_{\alpha}$ 
for all $T>0.46\epsilon$.
For $T < 0.46\epsilon$ the relaxation is already quite slow. 
At these temperatures, 
we used even longer initial relaxation times.
We also were using larger time-intervals 
between the saved configurations.
We provide more details on these 
technical details in the SM.
Thus, 
we estimate that 
for temperatures $T>0.42\epsilon$ the structures that
were used for data analysis can be considered 
as independent and well-relaxed.
At even lower temperatures, 
while we still were using the relaxation procedures,
the structures that were used for analysis cannot 
be considered as strictly independent.
Actually, to have structures which are more independent 
at low temperatures we were producing the structures 
from 10 independent runs 
(with 100 structures in each run for 8000 particles system).
The above details, 
expressed in terms of the $\tau_{\alpha}$ 
can also be reformulated
approximately in terms of the mean square displacement (MSD). 
Thus,
for $T>0.46\epsilon$, 
the MSD, 
$<(\Delta r)^2>$, 
which corresponded to 
the time interval between 
the two consecutively saved structures 
was always larger than $0.1\sigma^2$ for the $A$-particles.
See Fig.~\ref{fig:tcc-lll-diffusion}(a).

In summary, 
we believe that 
for $T>0.46\epsilon$ 
all saved structures 
can be considered as independent.
Then, 
in the temperature interval $0.42\epsilon \leq T \leq 0.46\epsilon$, 
the saved structures, 
obtained in the same simulation run, 
can also be considered as approximately independent.
For $T \leq 0.42\epsilon$ 
it is reasonable to assume that 
the structures 
obtained in the same simulation run
are not independent.
In any case, 
at all studied temperatures, 
we analyzed all of the obtained structures 
(1000 for the systems of 8000 particles)/(120 for the system of $64 \cdot 10^3$ particles), 
as independent.
The ISs were produced using the FIRE  algorithm 
\cite{bitzek2006structural,guenole2020assessment}
within the LAMMPS program 
Ref.\cite{Plimpton1995,thompson2022lammps,lammps}.
The criterion for the convergence of the minimization
process was: the change in the total potential energy in 
the current MD step has to be smaller than $10^{-4}$ of the energy
value. A maximum number of $10^{4}$ steps was allowed.
The value of the time step used was $dt = 0.002\tau$. 
The recommended time step for the FIRE algorithm is the
same as the time step used in MD simulations \cite{lammps}.
More details can be found in the SM.

\section{Definition of collineations \label{sec:chains-def}}
We say that 
there is a link or a bond 
between two large $A$-particles 
if the distance between them 
is smaller than the cutoff distance, 
$r_{AAc}$, 
corresponding to 
the position of the 1st minimum in 
the partial PDF of $AA$ 
particles, 
i.e., $r_{AAc}=1.43\sigma$.
We say that $N$ large $A$-particles form 
a $C(N,\alpha_{max})$ collineation if all 
nearest links in the collineation 
intersect at angles $\alpha > 180^{\circ} - \alpha_{max}$. 
In the SM, 
we describe the algorithm that 
has been used to find 
the collineations in the system.
%

\begin{figure*}
\begin{center}
\includegraphics[angle=0,width=6.6in]{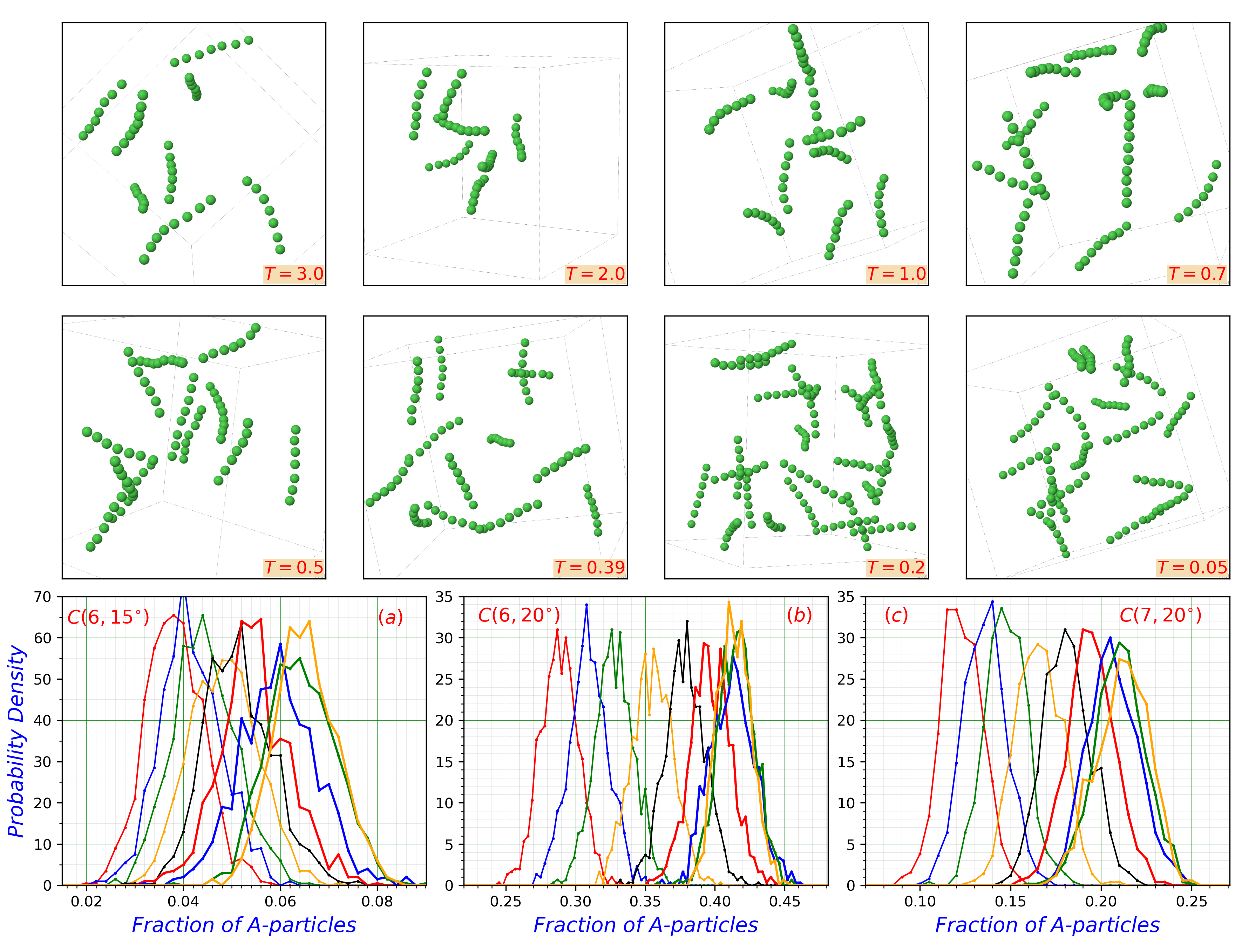}
\caption{
In a particular configuration of particles, 
at a given temperature, 
there are $A$-particles that form 
$C(7,15^{\circ})$ collineations. 
In two upper rows, 
only such $A$-particles are shown.
Different panels show 
the results from 
selected configurations 
at selected temperatures.
In the SM 
we provide several structure files, 
containing the coordinates of 
$A$-particles forming the collineations 
that can be examined with practically any
structure-view program.
In the lower row, 
we show 
the probability distributions for 
the numbers of $A$-particles 
forming different collineations 
at different temperatures. 
Panels $(a)$, $(b)$, and $(c)$ 
correspond to the collineations 
$C(6,15^{\circ})$, $C(6,20^{\circ})$, and
$C(7,20^{\circ})$ correspondingly. 
The abscissa-axes in these plots 
show 
the fraction of $A$-particles 
in the collineations 
relative 
to all $A$-particles in the system.
Different curves -- 
from the left to the right -- 
correspond to the temperatures 
$T = 3.0,\; 2.0,\; 1.5,\; 1.0,\;
0.7,\; 0.5,\; 0.39,\;
0.20,\; 0.05$.
}\label{fig:NC-prob-distr}
\end{center}
\end{figure*}

\begin{figure*}
\begin{center}
\includegraphics[angle=0,width=6.0in]{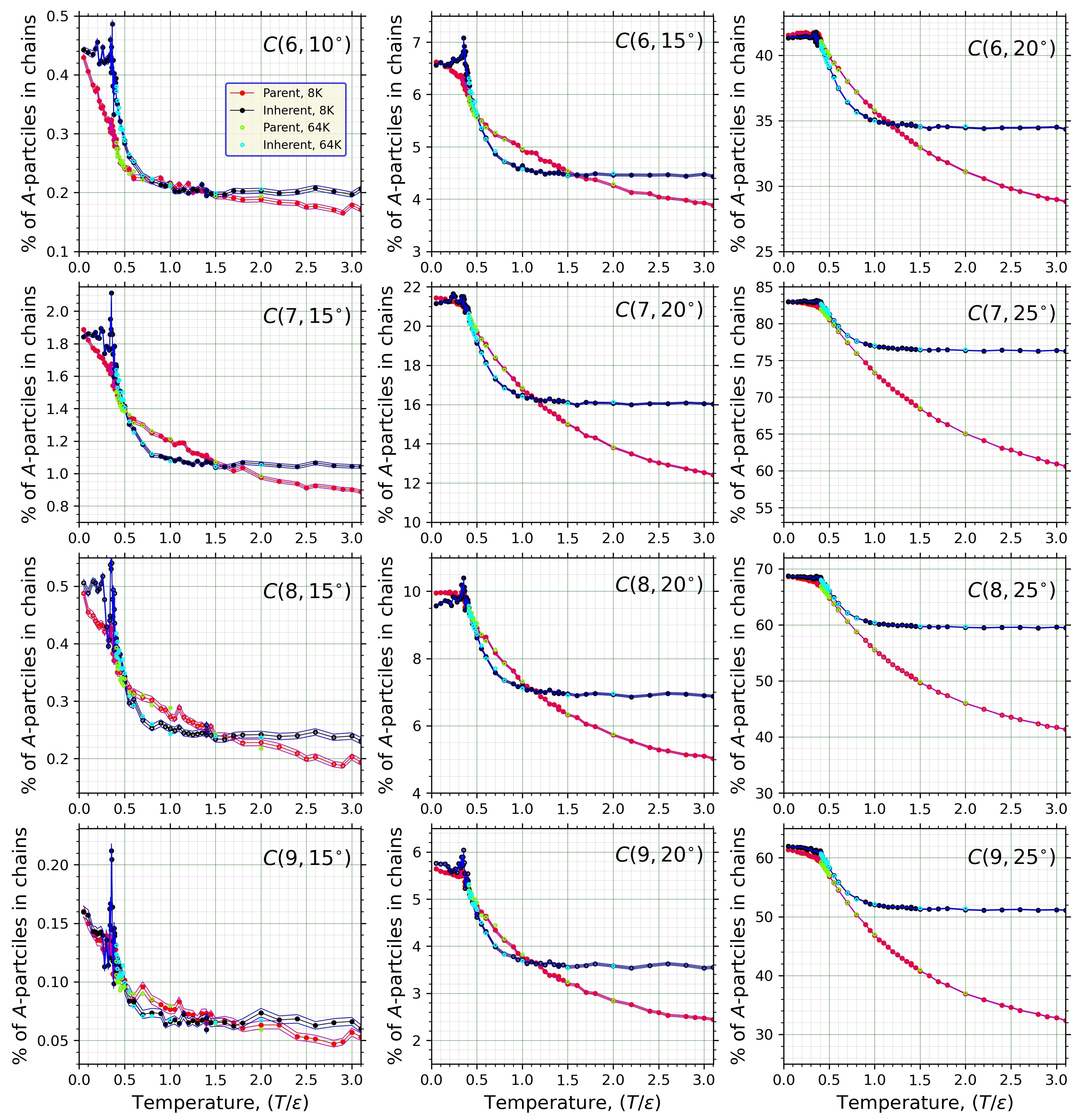}
\caption{
The dependencies of the average percentage of $A$-particles in all collineations of a particular type on the system's temperature.
The percentage is relative to the total number of $A$-particles in the system.
The panels in the same row 
correspond to the collineations 
with the same number of particles, 
but 
the limiting inclusion angles 
are different in different columns.
The red solid circles connected by the red lines show the results from the parent structures obtained on the system of 8000 particles. 
There are 
the average curves and also 
the average curves plus/minus 
the sigma of the mean (magenta).
The light green open circles 
correspond to 
the results obtained on 
the system of 
64000 particles. 
The error bars are not shown.
The black solid circles correspond 
to the results obtained on the inherent structures on the system of 8000 particles. 
The blue curves show the errors of the mean for the inherent structures. 
The cyan open circles correspond 
to the results obtained 
on the system of 64000 particles.
The results from the systems 
of 8000 and 64000 particles 
were obtained from 1000 and 120 
configurations correspondingly.
}\label{fig:NC-ave-vs-T}
\end{center}
\end{figure*}

\section{Observation of  collineations on the parent and inherent structures \label{ssec:observation-of-chains}}
Two upper rows 
in Fig.\,\ref{fig:NC-prob-distr} 
show the snapshots of
$C(7,15^{\circ})$ collineations 
formed by $A$-particles 
in the $8000$-particles system 
at selected temperatures. 
Only particles forming 
the $C(7,15^{\circ})$ 
collineations are shown.
It is clear that statistically
at lower temperatures 
there are more collineations than 
at higher temperatures.
We note that 
it does not follow 
from the shown data 
that the collineations at lower
temperatures agglomerate into 
some more complex structures.
Panels (a,b,c) in the 3rd row of 
Fig.\,\ref{fig:NC-prob-distr}
show the probability distributions 
for the numbers of particles 
involved in all collineations 
of a particular type 
in instantaneous configurations.
Here, in the counting procedure,
we adopted the rule that if some particle belongs 
to more than one collineation 
it is counted only once.
As the temperature decreases the probability distributions shift to the right. 
Thus, the numbers of particles in the collineations statistically increase. 
The shown data originate from 
1000 instantaneous configurations 
of 8000 particles produced 
in 10 independent runs.
In Fig.\,\ref{fig:NC-ave-vs-T}, 
the red curves and 
the red solid circles 
show how 
the average numbers of particles 
in all collineations of 
a particular type 
depend on the temperature.
In every panel, 
there are three red curves 
that show the average value and 
the average value $\pm \sigma_{mean}$.   
We see that the average numbers of particles 
involved in the collineations monotonically increase, 
as the temperature of the liquid decreases.  
There is a more abrupt rise in the numbers of collineations 
in the temperature range $0.3 < T <0.5$. 
In view of Ref.~\cite{pedersen2018phase,ingebrigtsen2019crystallization}, 
this behavior might be related to the proximity 
of the temperatures at which the crystallization process has been observed 
$T \sim 0.41 \pm 0.02$ 
\cite{ingebrigtsen2019crystallization}.
In the considered system, 
the crystallization process manifests itself, 
by the appearance of the regions 
containing FCC-crystals of $A$ particles.
At this stage, we will not address the origin of an abrupt rise in the number of collineations at $T \sim 0.41 \pm 0.02$ in more details.
In general, the hypothesis that collineations might represent a feature that is related to the formation of the crystal nuclei deserves a consideration.
The blue curves 
in Fig.\,\ref{fig:NC-ave-vs-T} 
correspond to the results 
obtained on 
the inherent structures (ISs). 
Practically, 
all shown blue curves 
exhibit an abrupt crossover 
at the potential energy landscape (PEL) 
crossover (PELC) temperature (PELCT). 
Originally, 
the PELCT was observed by consideration 
of the potential energies of the ISs
\cite{1998SastryPEL}.
As far as we know, 
there is still no complete understanding 
of the structural changes that
underlie the PELC. 
For example, in Ref.\cite{pedersen2010geometry} it has been shown that the instantaneous values of the inherent potential energy are related to the number of the Frank-Kasper bonds (major ligands) in the system.
On the other hand, 
as we demonstrate below, 
there is no a clear relation between the collineations and the major ligands. 
Thus, there appear to be several structural changes that can be connected to the PELC.
Another example of this situation is associated 
with the temperature dependence of the number of some clusters that have been introduced in the 
topological cluster classification 
approach \cite{malins2013identification,
malins2013lifetimes,
malins2013longlivedclusters}.
We will discuss this situation below.
Of particular interest is 
the relation between 
the red and blue curves
corresponding to the collineations 
$C(6,15^{\circ})$, $C(6,20^{\circ})$, 
$C(7,15^{\circ})$, $C(7,20^{\circ})$,
$C(8,15^{\circ})$, $C(8,20^{\circ})$, and 
$C(9,20^{\circ})$.
For these collineations, 
in the supercooled liquid range of temperatures,  
$0.4 \lesssim T \lesssim 1.0$, 
we observe that the average number of collineations
in the ISs is smaller than 
the number of collineations 
in the parent structures (PS). 
This is a counter-intuitive behavior. 
Indeed, since the ISs are more relaxed than the PSs, 
it is natural to expect that there always should be 
more collineations in the ISs.
In our view, 
the obtained results suggest that 
in the parent supercooled liquid 
there might develop some delicate, 
beyond medium-range, order 
that is destroyed
by a ``rough" relaxation procedure 
by which the ISs are produced from the PSs. 
The linear collineations 
should be related to this order.
It is also possible that the decrease 
in the number of collineations in the transition 
from the PS to IS is 
related to the fact that the pressure 
of the IS becomes 
negative when the temperature of the parent structures
is reduced below $T\approx 0.8$. 
The pressure of the IS becomes
even more negative as the temperature 
of the PS further decreases.
See the SM and, for example,  
Ref.\cite{
sastry2000liquid,
altabet2018cavitation,
makeev2018distributions}.
We note that in 
Ref.\cite{sastry2000liquid} 
it has been shown that
at $\rho_o = 1.2$, 
that we study in this paper, 
the ISs (under the negative pressure) 
remain globally homogeneous.
However, it is possible that in the process of relaxation from the PS into the IS in the system develop some local heterogeneities which can lead to the decrease in the number of collineation in the IS in comparison to the PS.
%

\begin{figure*}
\begin{center}
\includegraphics[angle=0,width=7.0in]{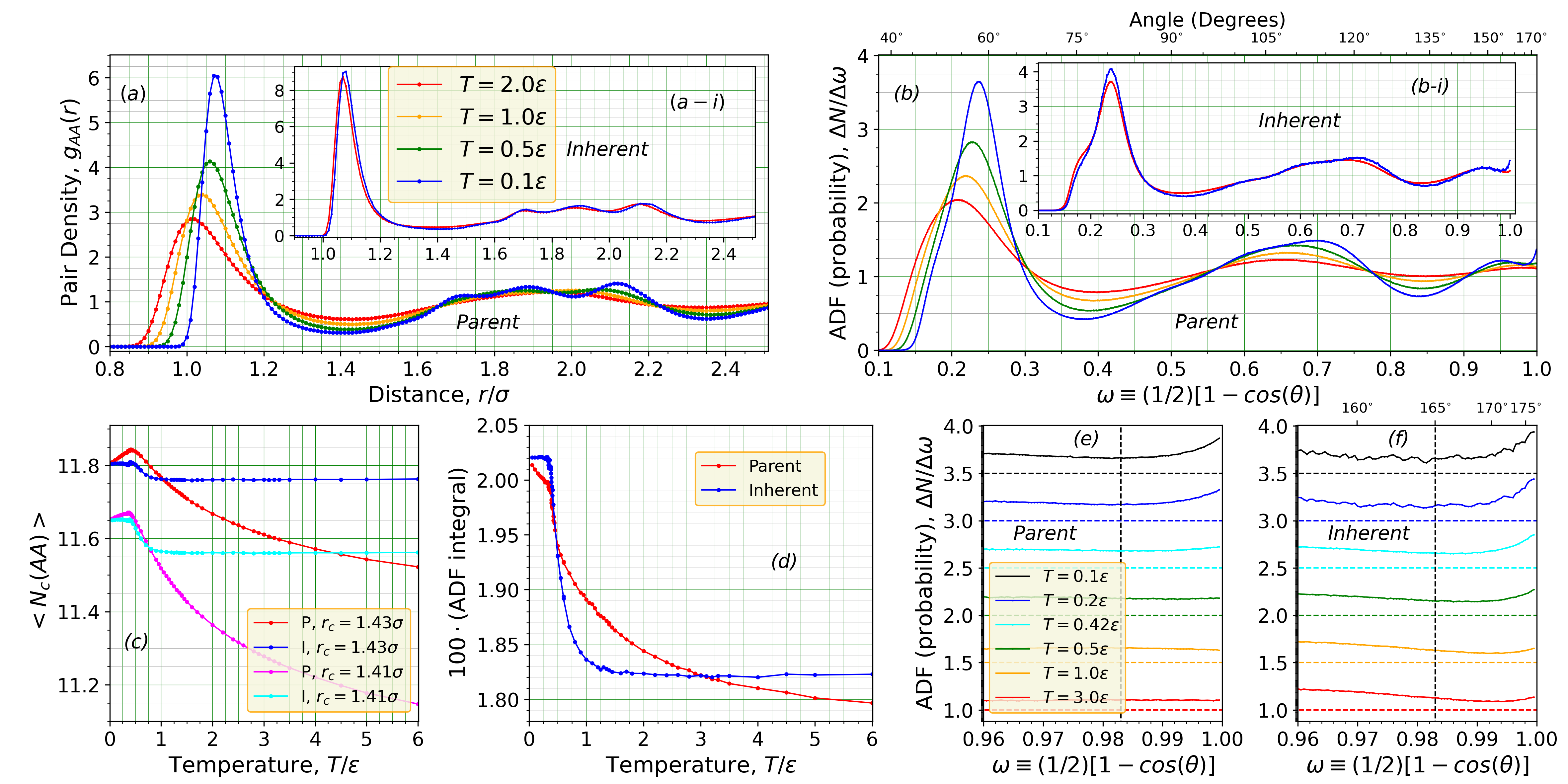}
\caption{
(a) Partial radial distribution function, $g_{AA}(r)$, 
for $AA$ particles at temperatures shown in the inset (a-i). 
In the inset, the $g_{AA}(r)$ calculated on the inherent structures are shown.
Note that the first minimum of $g_{AA}(r)$ is located at $r \approx 1.43\sigma$.
(b) The angular density functions (ADFs)
for the triplets of $A$ particles which are nearest neighbors 
at the same temperatures as in panel (a). 
In the inset (b-i), the ADFs calculated on the inherent structures are shown.
The temperatures of the corresponding parent structures are the same as in (a) and (a-i).
We are interested in the behaviors of the ADFs at $\omega \approx 1$, i.e., at $\theta \approx 180^{\circ}$. Thus, in panels (e) and (f) we show the ADFs in the vicinity of $\omega \approx 1$ for the parent and inherent structures correspondingly.
In panels (e,f), each curve corresponding to the next lower temperature is shifted upward by
0.5 with respect to the previous higher temperature for clarity.
The dashed horizontal lines correspond to the hypothetical homogeneous distribution of the $\bar{N}_c$ nearest neighbor particles over the surface of the sphere (if the normalization is to unity).
In panel (c), we show how the number of $A$-particles within the first minimum 
of $g_{AA}(r)$ depends on the temperature. 
The results for two different cutoff values of $r_c$ are shown.
In the legends of (c), the letters ``P" and ``I" correspond to the results 
obtained on the parent and inherent structures.
Panel (d) shows how the integral of the ADF over the range $\omega \in [0.983,1)$ depends on the temperature. In the plot, the values of these integrals were multiplied by 100. The results for the parent and inherent ADFs are shown. 
}\label{fig:PPDF-and-PADF}
\end{center}
\end{figure*}

\section{Analysis of the partial pair density and angular distribution functions \label{sec:Analysis-of-PDF-and-ADF}}

We use the following probabilistic model 
to connect the behaviors of the PDF and ADF, 
as the functions of temperature, 
with the numbers of particles in the collineations.
We assume that 
the number of nearest neighbors, $\bar{N}_c$, 
is the same for all $A$-particles.
We also assume that $\bar{N}_c$ depends on the system's temperature 
and 
that it corresponds to 
the integral over the first peak of 
the partial PDF (PPDF) (until the first minimum).
In Fig.~\ref{fig:PPDF-and-PADF}(a), 
we show the $AA$-PPDFs
for selected temperatures (shown in the inset).
In the inset, 
we show the PPDF calculated on the inherent structures.
The cutoff distance $r=1.43\sigma$ was used 
in the calculations of $\bar{N}_c$.
The curves in Fig.~\ref{fig:PPDF-and-PADF}(a,b) 
were obtained from averaging over 10 independent runs with 
100 structures in each run. 
For $T>0.46$ all structures in each run are separated by time intervals
which are several times larger than the $\alpha$-relaxation time.
Further, we assume that the ADF for nearest neighbors is also known
and that it depends on the temperature.
It is convenient to use the ADF 
in the form suggested by G.D. Scott et. al. 
\cite{scott1964angular,bernal1964bakerian},
i.e., consider the ADF as a function of the parameter
$\omega \equiv (1/2)[1-\cos(\theta)]$, where $\theta$ is 
the angle between two bonds that start on a chosen particle.
With this choice of the argument to the same intervals of 
$\Delta \omega$ correspond the same areas 
of the surface formed by the nearest neighbors. 
In Fig.~\ref{fig:PPDF-and-PADF}(b)
we show several $AAA$-partial ADFs (PADF) at selected temperatures.
In the inset, the PADFs calculated on two sets of inherent structures are shown.
All curves were calculated on the same structures that were used for producing
Fig.~\ref{fig:PPDF-and-PADF}(a).
The shown PADFs are normalized to unity.
Thus, 
if there were only one particle 
in the first-neighbor shell which were 
homogeneously distributed over this shell 
then the PADF would be equal to unity
in the whole range of $\omega$. 
According to the considered model, 
there are $\bar{N}_{c}(T)$ particles in the shell. 
Thus, the actual PADFs for these $\bar{N}_{c}(T)$ 
particles
are $\bar{N}_{c}(T)$ times larger than those shown 
in Fig.~\ref{fig:PPDF-and-PADF}(b).
Further, we focus on those collineations
in which all angles $\theta$
between the collineations-forming bonds 
are in the interval
$[180^{\circ} - 15^{\circ}; 180^{\circ}]$.  
This interval approximately corresponds
to the interval $\omega \in [0.983;1.0]$.
In Fig.\ref{fig:PPDF-and-PADF}(e,f),
we show the PADFs for 
the parent and inherent structures on a larger scale
which includes this interval of $\omega$. 
This interval is on the right 
with respect to the vertical dashed line.
Note, in Fig.\ref{fig:PPDF-and-PADF}(e,f), 
that
all curves obtained from the simulations are above 
the horizontal dashed lines
corresponding to the case of homogeneous distribution of neighbors. 
Note also that in the cases of 
parent and inherent structures, 
the separations between the curves 
from the simulations 
and 
the corresponding horizontal dashed lines 
increase as the temperature decreases.

\begin{figure*}
\begin{center}
\includegraphics[angle=0,width=7.0in]
{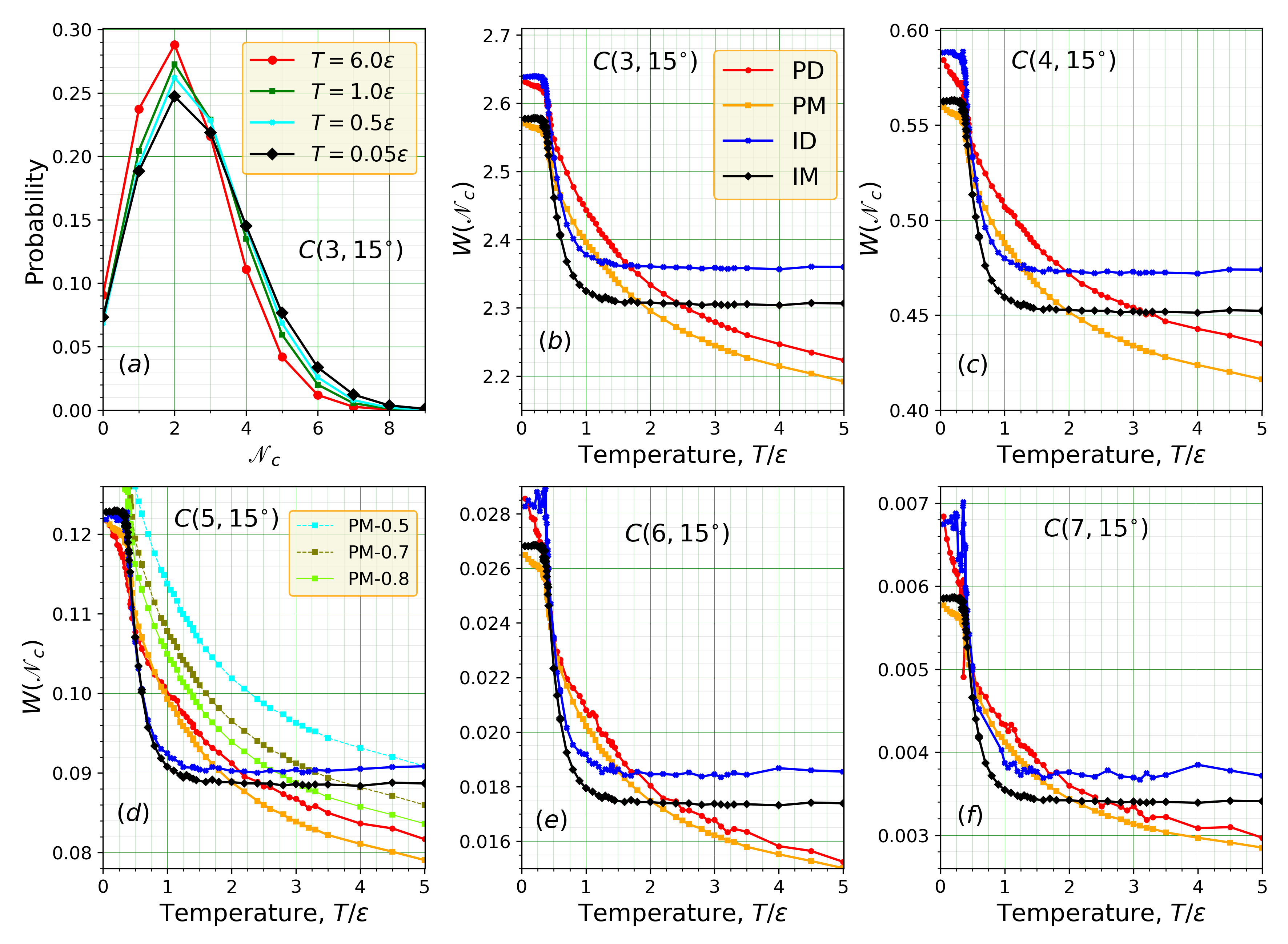}
\caption{
The temperature dependencies of the percentage of the number of particles located at the ends of some collineations. The percentage is with respect to the total number of $A$-particles in the system. In the panels, the results of direct calculations on the parent and inherent structures are shown together with the results from the model based on the analysis of the PDFs and ADFs. 
In the legends of panel (b), 
PD-stands for ``Parent Direct", 
PM-stands for ``Parent Model", 
ID-stands for ``Inherent Direct", 
and 
IM-stands for ``Inherent Model".
These notations are related to panels (b,c,d,e,f).
In the legends of panel (d) the notation
``PM-0.5" corresponds to the model curve in calculations of which we used $(\bar{N}_c - 0.5)$ in Eq.(\ref{eq:Nc-prob}) instead of $(\bar{N}_c - 1)$.
Note that in certain ranges of temperatures there are more particles forming the collineations in the parent structures than in the inherent structures.
}\label{fig:Chains-Direct-vs-Model}
\end{center}
\end{figure*}

The most interesting feature of the shown PADF curves 
arises when the liquid on cooling turns 
into a glass (approximately when $T < 0.40$). 
In the glassy state, 
there develops a pronounced 
increase of the PADF 
towards $\omega = 1$, i.e., 
towards $\theta = 180^{\circ}$.   
This increase is seen in the interval $ 0.994 < \omega \leq 1$. 
In the PADFs obtained on the IS 
this increase towards $\omega = 1$ 
is present for all temperatures of the original PS. 
However, the increase is larger for the IS obtained 
from the PS at lower temperatures 
and 
especially from the PS 
which are below the GT temperature.
This behavior, in our view, 
indicates that the rise
of the PADF towards $\omega = 1$ is associated with
development of a structural organization in 
the low-temperature supercooled liquid and the glass states
which is not properly
captured by the quick relaxation procedures by which the IS are 
obtained from the PS. 
This increase poses two questions: 
1) Can this increase be considered as a characteristic structural 
feature of the glassy state?
2) Is the origin of this effect local or non-local?
We will discuss this issue in the context of 
Fig.~\ref{fig:Chains-Direct-vs-Model}(b,c,d,e,f).

It is obvious that the increase in  the ADF at
$\theta \approx 180^{\circ}$ is related 
to the development of the splitting 
of the second peak in the PDF. 
In particular, to the observation 
that this peak is split in the glass state, 
but not in the liquid state 
\cite{royall2015role,cheng2011atomic,
finney1970random, bennett1972serially, 
wendt1978empirical, voloshin1997origin, 
truskett1998structural,o2005structure,
luo2006pair,pan2011origin,ding2015second,wu2015hidden}.

Further, 
we assume that
the probability for a 
chosen central $A$-particle and one of its $A$-neighbors 
that there is another $A$-neighbor 
which is located approximately opposite to the first neighbor is: 
\begin{eqnarray}
p_3 \equiv (\bar{N}_{c}-1)p_{\omega} \equiv (\bar{N}_{c}-1) \int_{\omega = 0.983}^{\omega = 1} f_{PADF}(\omega)d\omega.\;\;\;
\label{eq:p3}
\end{eqnarray}
In (\ref{eq:p3}), 
we put $(\bar{N}_c-1)$ in front of the integral because we consider the probability for the second
neighbor to be located oppositely enough to the first neighbor which was already chosen. 
The notation $p_3$ is used because (\ref{eq:p3}) addresses the case of three particles.

In Fig.\ref{fig:PPDF-and-PADF}(d), 
we show the temperature dependence of the integral from 
(\ref{eq:p3}) (without the $(\bar{N}_{c}-1)$ prefactor).
The first thing to notice is that the value of this integral calculated on the IS clearly indicates existence 
of the crossover temperature
$T_{x}$. 
Note that here, contrary 
to the standard considerations \cite{1998SastryPEL,debenedetti2001supercooled},
we see this behavior in a quantity which is directly related to a particular structural property.
In the standard considerations, it is not clear what structural changes
cause the crossover at $T_{x}$ in the PEL \cite{1998SastryPEL,debenedetti2001supercooled}.

Another feature to notice in Fig.\ref{fig:PPDF-and-PADF}(d)
is that in the interval of temperatures $0.5 < T < 3.0$ 
the red curve calculated on the
PS is above the blue curve calculated on the IS. 
This again indicates, in our view, 
that in the liquid there develops some structural order
that is being destroyed by the quick relaxation 
procedure leading to the IS.

In Fig.\ref{fig:PPDF-and-PADF}(c), 
we show, 
for the PS and IS, 
how the average $A$-partial coordination number of $A$-particles, 
depends on the temperature for two values of the cutoff distance.
The curves in 
Fig.\ref{fig:PPDF-and-PADF}(c) 
were obtained
by integration of the curves like those in Fig.\ref{fig:PPDF-and-PADF}(a).
It is of interest that the curves calculated on the PS clearly indicate 
the GT temperature while the curves calculated on the IS 
in addition exhibit the crossover temperature $T_x$.

\section{Connection of the PDF and ADF with the number of particles in collineations \label{sec:PDF-ADF-Chains}}

In the discussions of collineations,
the most natural question to ask is what is 
the number of particles 
in the collineations of a particular type? 
We addressed this question in 
Fig.\ref{fig:NC-prob-distr},\ref{fig:NC-ave-vs-T}.
However, to understand the connection between the PDF, ADF, 
and collineations it is more convenient to ask a different 
closely related question. 
That is: 
What is the probability for a randomly chosen $A$-particle to be located at the end (start) of 
any collineation with $\mathscr{N}_c$ particles in it? 
Actually, the answer is quite obvious:
\begin{eqnarray}
W(\mathscr{N}_c)=\bar{N}_c \cdot \left[(\bar{N}_c -1)p_{\omega}\right]^{\bar{N}_c -2}.
\label{eq:Nc-prob}
\end{eqnarray}
In (\ref{eq:Nc-prob}), the first $\bar{N}_c$ on the rhs is the number of directions
in which the collineation can be directed if it starts on a chosen particle. 
Then, $\left[(\bar{N}_c -1)p_{\omega}\right]$ with $p_{\omega}$ from (\ref{eq:p3}) is the probability
that there is a third particle in the collineation that is located at the proper angle with respect to the bond that
joins the chosen start particle and its already chosen neighbor, i.e., the second particle in the collineation.
Then, another factor $\left[(\bar{N}_c -1)p_{\omega}\right]$ comes from the second angle in the collineation.
Since there are $(N_c - 2)$ angles in the collineation of $\bar{N}_c$ particles we arrive to the expression (\ref{eq:Nc-prob}). Of course, we assumed that the probabilities for all
angles are independent.

Note that $W(\mathscr{N}_c)$ can be larger than unity. 
In this case, $W(\mathscr{N}_c)$, effectively becomes equal to the
average number of collineations that start on a chosen start particle. 

In Fig.~\ref{fig:Chains-Direct-vs-Model}(b,c,d,e,f),
we show how $W(\mathscr{N}_c)$ 
depends on the temperature 
for the collineations with different
$\mathscr{N}_c$  (3,4,5,6,7). 
In Fig.~\ref{fig:Chains-Direct-vs-Model}(b,c,d,e,f), 
the red curves show the results
directly calculated on the parent structures.
The orange curves show the results calculated according
to Eq.~(\ref{eq:p3},\ref{eq:Nc-prob}). 
Again, we note that in these calculations 
we take into account
that $N_c$ and $p_{\omega}$ depend on the temperature through
the temperature dependencies of the PDF and ADF, 
as shown in Fig.~\ref{fig:PPDF-and-PADF}.
Similarly, the blue curves 
in Fig.~\ref{fig:Chains-Direct-vs-Model}(b,c,d,e,f) 
show the results directly calculated
on the inherent structure, while the black curves
show the results derived from the PDFs and ADFs
according to Eq. (\ref{eq:p3},\ref{eq:Nc-prob}).

Note in Fig.~\ref{fig:Chains-Direct-vs-Model}(b) 
that $W(\mathscr{N}_c)$ for the particles
in $C(3,15^{\circ})$ collineations
is larger than two. 
Thus, in Fig.~\ref{fig:Chains-Direct-vs-Model}(a) 
we plot,
for the parent structures, 
the probability distributions for the particles to have
different $\mathscr{N}_c$ 
(for the $C(3,15^{\circ})$ collineations).
These results were obtained from the direct analysis of the structures.

In expression (\ref{eq:Nc-prob}), 
in the square brackets,
we have an expression $(\bar{N}_c -1)p_{\omega}$. 
We explained above why we use $(\bar{N}_c -1)$ 
instead of simply $\bar{N}_c$. 
Yet, it is of interest to have 
a  better understanding of how the choice of
$(\bar{N}_c -1)$ 
instead of 
$\bar{N}_c$ 
affects the curves in Fig.~\ref{fig:Chains-Direct-vs-Model}(b,c,d,e,f).
In Fig.~\ref{fig:Chains-Direct-vs-Model}(d), we address this point
by presenting also the curves for 
$(\bar{N}_c -0.8)$, 
$(\bar{N}_c -0.7)$, and 
$(\bar{N}_c -0.5)$. 
These curves are marked as 
$(PM-0.8)$, $(PM-0.7)$, and $(PM-0.5)$.
We see that the curves are quite sensitive to the choice of 
$(\bar{N}_c -1)$ instead of $\bar{N}_c$.

\section{Discussion of the possible connection with the Frank-Kasper and Geometrical Fustration approaches \label{sec:Discussion-Frank-Kasper}}

As we already mentioned in the introduction, 
the collineations may be related to several other phenomena previously  discussed in the context of supercooled liquids and glasses. 
One of these is the concept 
of major skeleton introduced by Frank and Kasper in 
Ref.~\cite{frank1958complex}. 
Another is the concept of ``disclination lines" that is important in the geometrical frustration approach ~\cite{kleman1979tentative,sadoc1982order,
nelson1983order,
qi1991icosahedral,derlet2020correlated}. 
These two concepts are closely related ~\cite{
nelson1983order,qi1991icosahedral,
pedersen2010geometry,derlet2020correlated}.
Investigation of a possible relation between 
the collineations and the major skeleton or the disclination 
lines will also allow us to address 
the local structure near the collineations.

In the basis of the Frank-Kasper and geometrical frustration approaches lies the observation that four particles can be ideally packed into the tetrahedron 
\cite{frank1958complex,kleman1979tentative,sadoc1982order,
nelson1983order,derlet2020correlated}.
Then, still from the local perspective, 
it is recognized that twenty tetrahedra, 
if slightly distorted, 
can be packed into the icosahedron 
which locally provides more dense 
packing than the FCC or HCP lattices. 
Also, since the surface edges of 
the icosahedron are $5\%$ longer 
than the distance from the center of icosahedron 
to its vertices, the packings into 
icosahedra provide more flexibility 
in the packings of particles ~\cite{frank1958complex}. 
However, it is impossible to fill 
the whole space with tetrahedra or icosahedra alone. 
Thus, it is necessary to understand 
the nature of defects in the packings of tetrahedra (or icosahedra). 

The nature of the defects in the FK and geometrical frustration approaches is analyzed from two different perspectives, though practical conclusions, in many respects, are similar. 

In the FK approach \cite{frank1958complex}, 
under the assumption that the distortions of the tetrahedra 
should be relatively small, 
it is argued that besides the particles at the centers of the icosahedron environments (``minor sites"), 
with coordination of the central particle being 12, 
there are also ``major sites" with coordinations of 
the central particle being 14, 15, 16, but not 13. 
The connection between the major sites is labeled
as a {\it{major ligand}}.
The number of major ligands for a particle 
with the coordination number $Z$ is $(Z-12)$.
Thus, the number of major ligands is 
2 for $Z=14$, 
3 for $Z=15$, and 
4 for $Z=16$.
Correspondingly, 
if there is one major site, 
then between its nearest neighbors there should be at least two other major sites. Thus, it was argued that the major sites form the ``major skeleton" of the structure. It was also concluded that two nearest major sites should have at least 6 neighbors in common, while nearest minor sites have 5 neighbors in common. The links between the major and minor sites also should have 5 neighbors in common.
Further, it was argued that, 
like in certain transition metals and alloys, 
the major skeleton should have a tendency to be linear, while minor sites, i.e., the sites with coordination 12 should have a tendency 
to be located in the pseudoplanes approximately orthogonal to the lines of the major skeleton. 

In the geometrical frustration approach \cite{kleman1979tentative,sadoc1982order,
nelson1983order,qi1991icosahedral,
derlet2020correlated}, 
the connectivity and geometry of defects (disclination lines) is introduced through initial considerations of the packing of hard particles on the three dimensional surface of a four dimensional sphere. This surface is a curved three-dimensional space of finite size. It was demonstrated that it is possible to pack 120 particles on this surface in a crystal-like 
structure in which every particle is surrounded by twelve other particles in the (curved) icosahedral geometry. Then, it is assumed that if this structure is ``projected" into 3D Euclidean space, in an energy efficient way, then the resulting structure is a good approximation to the structure of some glasses and liquids. In order to perform the operation of ``projection", it is necessary for some pairs of neighbor particles, that initially have 5 neighbors in common, to introduce an additional particle into the set of common neighbors. For other pairs it may be necessary to remove some of the common neighbors. In this way, in the system are introduced negative and positive disclinations lines. The particles that form negative/positive ($-72^{o}/+72^{o}$) disclinations lines should have coordinations more/less than twelve. Also,  the number of the common neighbors of the neighbor pairs of particles that form negative/positive disclination lines should be larger/smaller than five.

Thus, in order to address a possible relation between the collineations and the major skeleton or between the collineations and the disclinations lines it should be sufficient, at least as the first step,
to compare statistically the coordinations of the $A$-particles that form collineations with the coordinations of all $A$-particles \cite{qi1991icosahedral,derlet2020correlated}. 
It is also reasonable to compare statistically the number of the common neighbors for the pairs of $A$-particles that form collineations with all pairs of $A$-particles that are nearest neighbors.

\subsection{Nearest Neighbor Analysis \label{ssec:NN_analysis}}

In general, there are two major approaches for determining
neighbors of a particle. 
In one approach, 
the neighbors are defined according to a chosen cutoff distance(s)\cite{hansen2013theory,honeycutt1987molecular,stukowski2012structure,royall2015role}. 
In another approach, the neighbors are defined according to a Voronoi-type tessellation \cite{gellatly1982characterisation,stukowski2012structure,lazar2022voronoi}. 
In both of these approaches there arise ambiguities with respect to the neighbor counting \cite{stukowski2012structure,derlet2020correlated,wang2021inconsistency}.
Here, we present the results of the neighbor analysis performed in both ways.
 
In the chosen cutoff approach, 
the neighbor cutoff distance for 
the pairs of $A$ particles was $r_c(AA) = 1.43\sigma$ and
for the pairs of $A$ and $B$ particles 
the cutoff was $r_c(AB) = 1.25\sigma$. 
For the KA system, the shape of the partial
PDF for the pairs of $B$ particles is such that it is not
quite clear how to define two particles as neighbors \cite{1995KobAndersen01,1995KobAndersen02}.
For our present considerations, 
the case of two $B$-particles 
which are neighbors is irrelevant.

As the Voronoi-type analysis we used the modified radical 
Voronoi (modRV) tessellation as it has been described in
\cite{derlet2020correlated,
gellatly1982characterisation,
stukowski2009visualization,
rycroft2009voro++}.
The radical Voronoi tessellation can be used if the studied system
consists of particles of different sizes 
\cite{gellatly1982characterisation}.
The modification used in the approach eliminates 
from the consideration 
some short edges of 
the Voronoi cells ~\cite{derlet2020correlated}.
This leads to better results for
the number of common neighbors 
for some pairs of particles and
eliminates from the consideration 
some strongly distorted tetrahedra ~\cite{derlet2020correlated}.

Let $l_{AA}=1.43\sigma$ and $l_{AB}=1.25\sigma$ 
be the distances at which the first minimums 
in the partial $g_{AA}(r)$ and $g_{AB}(r)$ are located.
In our analysis with modRV ~\cite{derlet2020correlated}, 
we the used the following values for the radii 
of $A$ and $B$ particles 
$R_{A} = l_{AA}/2 = 0.715\sigma$ and  
$R_{B} = l_{AB} - R_{A}=0.535\sigma$.

\begin{figure*}
\begin{center}
\includegraphics[angle=0,width=7.0in]
{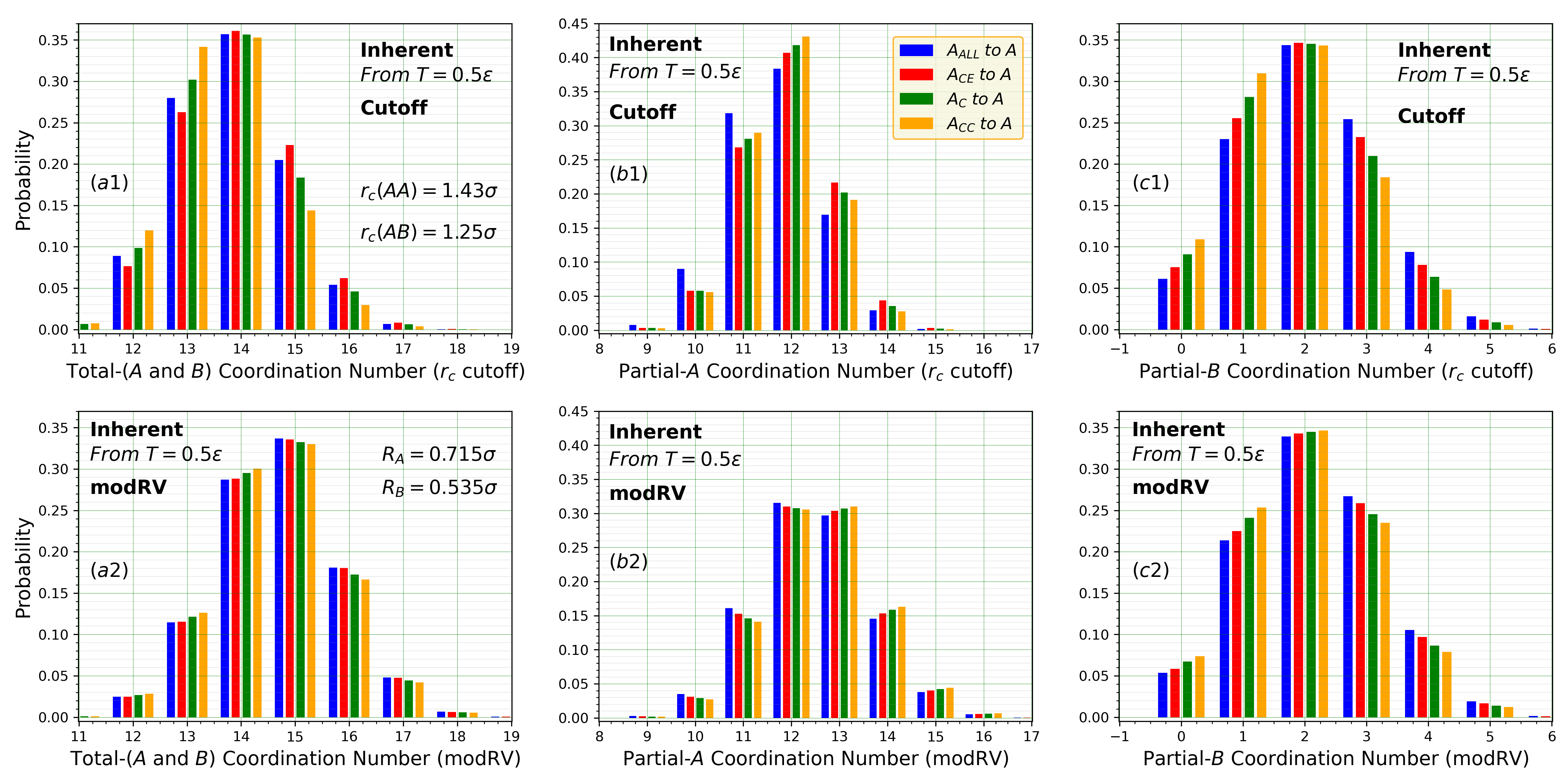}
\caption{
Results of statistical investigations
of local environments of $A$ particles 
in the $C(5,15^{\circ})$ collineations 
according to the chosen cutoff distances 
(a1,b1,c1) and 
according to 
the modified radical Voronoi (modRV)
tessellation 
(a2,b2,c2).
In panels (a1,a2), the results for the coordination due to $A$ and $B$ particles together
are shown. 
In panels (b1,b2)/(c1,c2), the results for the partial coordinations due to $A$/$B$ particles only are shown.
The blue/red/green/orange histograms 
(from the left to the right) 
marked in the legends of (b1)
as $A_{ALL}$/$A_{CE}$/$A_{C}$/$A_{CC}$ 
correspond to the data obtained on 
(all $A$ particles in the system)/(particles located at the ends of $C(5,15^{\circ})$ collineations)/(all particles in the $C(5,15^{\circ})$ collineations)/(from the triplets of particles in the centers of $C(5,15^{\circ})$ collineations). 
}\label{fig:Coordination-Number-CN-Cutoff-and-Voronoi}
\end{center}
\end{figure*}

\begin{figure*}
\begin{center}
\includegraphics[angle=0,width=7.0in]
{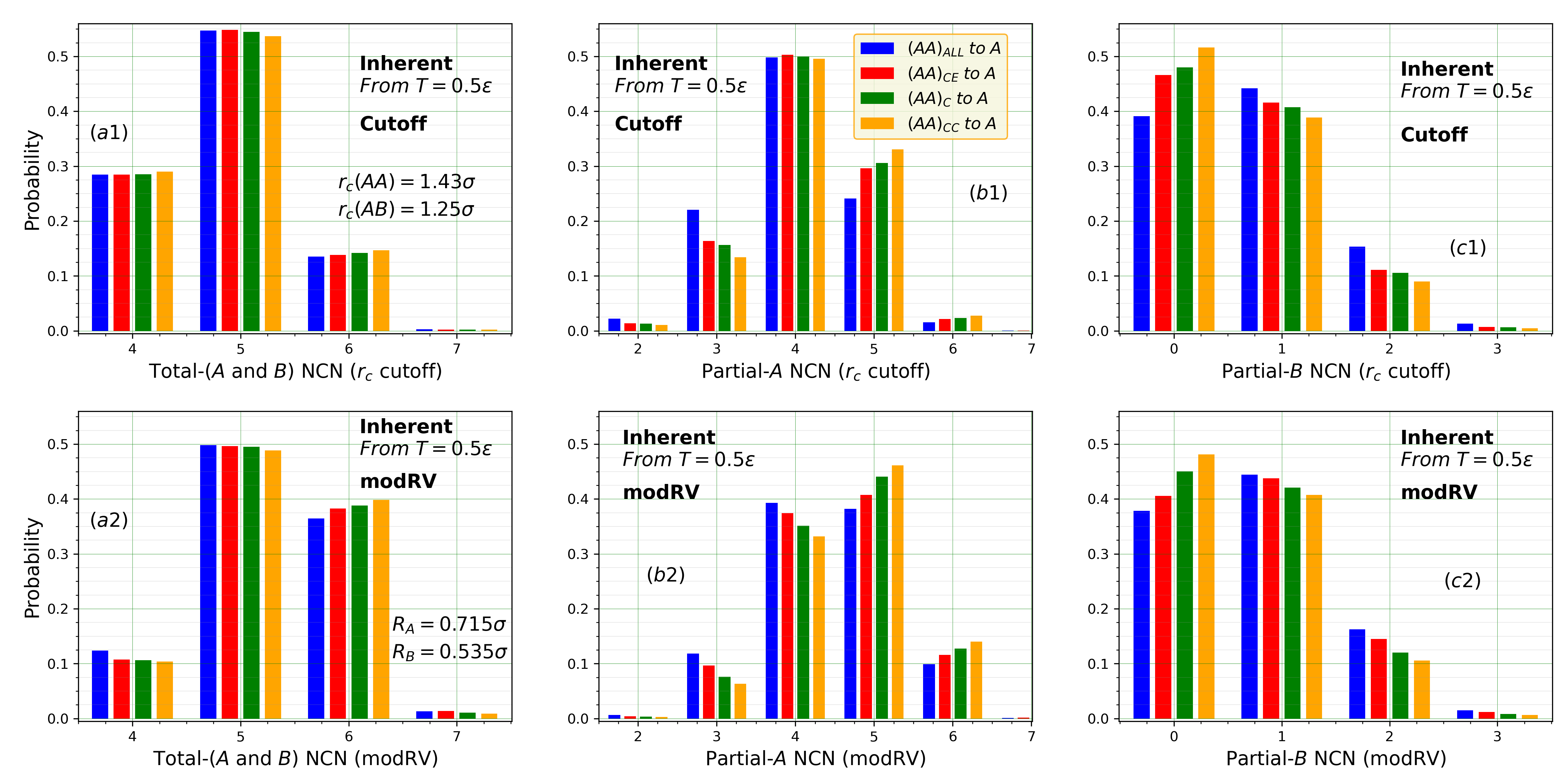} 
\caption{
Results of the common neighbors
analysis
for all $AA$-bonds and those
$AA$-bonds that form $C(5,15^{\circ})$ 
collineations (NCN-Number of Common Neighbors).
In panels (a1,b1,c1) the neighbors of each particle
are defined according to the chosen cutoff distances while in (a2,b2,c2) the neighbors are defined according to the modRV analysis.
In (a1,a2) we show the PDs for the number of common neighbors of the $AA$-pairs if the neighbors of both types are counted. In (b1,b2)/(c1,c2) only the common neighbors of type $A$/$B$ were counted.
We note that the neighbor particles in the collineations were always defined according to the
cutoff distance.
The most left (blue) bars in all panels correspond to the PDs calculated on all $AA$ bonds.
The second-left (red) bars were obtained from 
the analysis of neighbors of two $AA$ bonds that are 
located at the ends of the $C(5,15^{\circ})$ collineations.
Then, the third from the left (green) bars were obtained by averaging over 
all bonds in the collineations.
Finally,  the most right (orange) bars were obtained 
by averaging over two bonds in the collineations which 
are at the centers of the collineations. 
The legends in panel (b1) are for all panels.
Note that if a particular $C(5,15^{\circ})$ collineation belongs to some
longer collineation then some bonds in it can be counted in different roles.
In such cases, 
every count was considered as independent.
}\label{fig:Common-Neighbor-Analysis-CNA-Cutoff-and-Voronoi}
\end{center}
\end{figure*}

In Ref. \cite{qi1991icosahedral,derlet2020correlated}, 
the structural analysis of the inherent structures 
was performed to evaluate the applicability 
of the Frank-Kasper and geometric frustration considerations
to the studied system. 
For this reason, we also present in 
Fig.~\ref{fig:Coordination-Number-CN-Cutoff-and-Voronoi},
\ref{fig:Common-Neighbor-Analysis-CNA-Cutoff-and-Voronoi} 
the results of 
the nearest neighbor and common neighbor analyses 
performed on the inherent structures. 
However, we also analyzed the parent structures. 
The results from the parent structures are very similar to those in 
Fig.~ \ref{fig:Coordination-Number-CN-Cutoff-and-Voronoi},
\ref{fig:Common-Neighbor-Analysis-CNA-Cutoff-and-Voronoi}
not only from the qualitative perspective but also from the quantitative
point of view. 
The results from the parent structures are shown in
Fig.\,3,4 in the supplemental materials.

The results of the nearest neighbor analysis 
of the inherent structures produced from the parent structures 
at $T=0.5\epsilon$ 
are shown in 
Fig.~\ref{fig:Coordination-Number-CN-Cutoff-and-Voronoi}. 
In this figure, the histograms of different
colors correspond to different locations of the $A$ particles in the collineations 
(or not in collineations at all).

The first thing to note from 
Fig.~\ref{fig:Coordination-Number-CN-Cutoff-and-Voronoi}(a1,a2)
is that the coordination 
of the majority of $A$ particles is larger than 12.
This situation is markedly different from the situation 
in 
Ref.~\cite{qi1991icosahedral} (see, in particular, Table 1) or
in
Ref.~\cite{derlet2020correlated} (see, in particular, Fig. 3) where
a very significant fraction of particles had coordination number 12.
The presence in large amount of particles with icosahedral geometry 
and correspondingly with coordination number
12 is essentially a condition 
for the applicability of standard interpretations associated with 
the geometric frustration approach.
Indeed, in the geometric frustration approach, the particles with
coordinations larger or smaller than 12 are considered 
as excitations associated with the network of disclination lines Ref.~\cite{nelson1983order,qi1991icosahedral,derlet2020correlated}.
Thus, if the number of particles with coordination number 12 is small 
then it is not quite clear if the particles with coordinations larger 
or smaller than 12 can be considered as excitations.

We should note here that in 
Fig.~\ref{fig:Coordination-Number-CN-Cutoff-and-Voronoi}(a1,a2)
we considered as central only the particles of type $A$ 
which are larger than $B$-particles.
Since $B$-particles are smaller than $A$-particles, 
it may seem reasonable to assume that there will be $B$-particles which are the centers of the icosahedra.
However, it was shown in Ref.~\cite{coslovich2007understanding}
that in the KA system the nearest neighbor environments of the smaller $B$-particles 
practically always are not icosahedra.

Thus, 
in our view, 
it is not quite clear if 
the geometric considerations developed in the context of the 
Frank-Kasper and geometric frustration approaches 
are applicable in the case of the KA system.
The fundamental reason for this situation might be the following. 
The systems of particles considered in 
Ref.~\cite{qi1991icosahedral,derlet2020correlated} 
both crystallize into the Laves crystal 
structures \cite{qi1991icosahedral,pedersen2010geometry,derlet2020correlated}.
Thus, in agreement with the original considerations \cite{frank1958complex},
the tetrahedra and icosahedra are indeed favored local configurations for these systems.
The situation with the KA system is different.
From 
Ref.\cite{pedersen2018phase,coslovich2018dynamic,ingebrigtsen2019crystallization}, 
it is known that the KA system 
has a tendency 
for separation of the $A$ and $B$ particles 
in the process of crystallization. 
Then, the separated $A$-particles 
have a tendency to crystallize into the FCC structure.
Thus, it seems to us that the situation
with the KA system indeed might different in comparison with the systems
studied in 
\cite{qi1991icosahedral,pedersen2010geometry,derlet2020correlated}.
In any case, to address this issue in detail, 
it might be necessary to perform a more detailed study, 
like those in Ref.~\cite{qi1991icosahedral,derlet2020correlated}.

The comparison of 
Fig.~\ref{fig:Coordination-Number-CN-Cutoff-and-Voronoi}(a1) 
with 
Fig.~\ref{fig:Coordination-Number-CN-Cutoff-and-Voronoi}(a2) 
shows that the modRV method leads
to larger total coordination numbers than the cutoff approach. 
Further, the comparison of [(b1) with (b2)] and [(c1) with (c2)]
shows that the difference arises 
from how the pairs of $A$-particles are counted as neighbors. 
Indeed, the results in [(b1) and (b2)] are noticeably different.
On the other hand, the results for the pairs of $A$ and $B$ particles 
in [(c1) and (c2)] are quite similar.
This is a surprising result because there is some degree of ambiguity
in how to define the radius of the $B$-particles
in the modRV approach. 
Yet, the results in (c1) and (c2) are similar.

We now turn attention to the major question of interest.
Do particles forming collineations have environments different from other particles?
We see from Fig.~\ref{fig:Coordination-Number-CN-Cutoff-and-Voronoi} 
that different locations of the particles in collineations
weakly affect the probability distributions. 
Note also that the probability distributions for the particles in collineations are not that
different
from the probability distribution for all $A$-particles. 
The largest difference can be observed in 
Fig.~\ref{fig:Coordination-Number-CN-Cutoff-and-Voronoi}(b1). 
In general, it follows from the distributions that 
particles that belong to the collineations tend to have 
larger numbers of $A$-neighbors 
and 
smaller numbers of $B$-neighbors.
This effect can be considered as a possible indication of the relation
of collineations to the tendency of the system for components separation.

The general conclusion from 
Fig.~\ref{fig:Coordination-Number-CN-Cutoff-and-Voronoi}
is that statistically the average environments of the $A$-particles
forming the collineations are slightly different than the average environment
of all $A$-particles. 
Thus, it is possible to say that statistically the environments 
of the $A$-particles forming the collineations 
exhibit some tendency for some specific orderings.
However, we clearly cannot claim from these results that the $A$-particles
forming the collineations have some fixed geometry.

\subsection{Common Neighbor Analysis \label{ssec:CNA_analysis}}

Another way to address the differences in the environments 
of the particles forming the collineations and other particles is 
to perform the common neighbor analysis (CNA) 
for the pairs of the nearest neighbors particles \cite{honeycutt1987molecular,jonsson1988icosahedral}. 
Thus, 
in Fig.~\ref{fig:Common-Neighbor-Analysis-CNA-Cutoff-and-Voronoi},
we show the results of the CNA for the four sets of particles
as described in the caption of the figure.

The results in 
Fig.~\ref{fig:Common-Neighbor-Analysis-CNA-Cutoff-and-Voronoi}(a1,a2)
show that the majority of the bonds are the 5-fold bonds,
supporting the point of view that the icosahedral ordering
plays very important role in supercooled liquids and glasses \cite{frank1958complex,
kleman1979tentative,sadoc1982order,
nelson1983order,
qi1991icosahedral,
derlet2020correlated,
jonsson1988icosahedral,tarjus2005frustration}.
On the other hand, the total amount of the 4-fold and 6-fold bonds together
is approximately the same as the number of 5-fold bonds.
Thus, in our view, it is not quite clear if these 4-fold and 6-fold bonds
should be considered as excitations.
It is of interest that in panel (a1) the number of 4-fold bonds is larger than
the number of 6-fold bonds while in panel (a2) the number of 6-fold bonds
is larger than the number of 4-fold bonds. 
In the frame of the geometric frustration approach, 
in Euclidean flat space the number of 6-fold bonds 
should be larger than the number of 4-fold bonds \cite{nelson1983order}. 

As far as it concerns the dependence of the bonds' coordination on the
location of the bonds in collineations, 
we do not observe a significant dependence in panels (a1,a2).
This dependence is more pronounced in (b1,b2) and in (c1,c2). 
The differences have a statistical character. 
The major tendency that we observe is that bonds in the collineations
tend to have larger 
number of $A$-neighbors and smaller number of $B$-neighbors.
This situation is in agreement with the results from the 
analysis of the coordination number.

We summarize our results
with respect to the relation of collineations to the
disclinations lines as follows. 
If it is possible to speak about the disclination lines
in the KA system, 
then it is unlikely that there is a connection between 
the collineations and the disclination lines.
%

%
\begin{figure*}
\begin{center}
\includegraphics[angle=0,width=6.5in]
{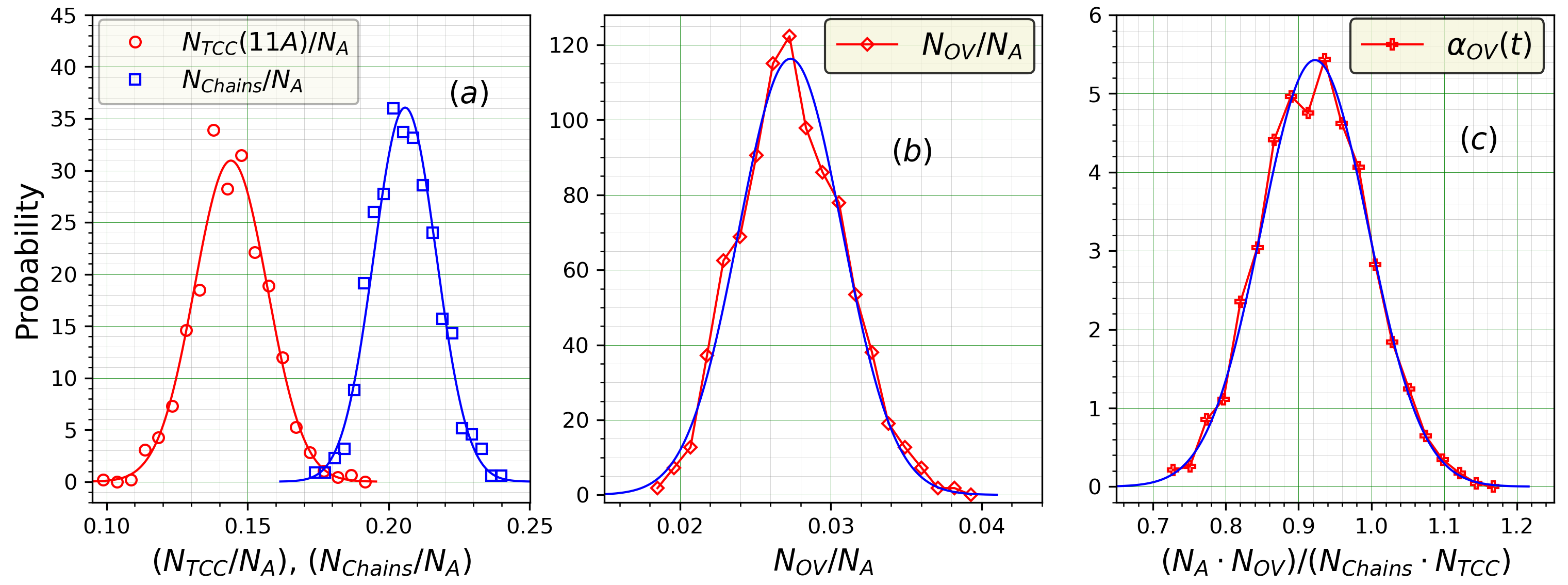}
\caption{
Addressing the non-randomness of the overlap between the $C(5,15^{\circ})$ collineations 
and the $11A$-TCC clusters at $T=0.5$.
(a) The probability distributions (PDs) 
for the total numbers
of particles (as the fractions of $N_{A}$) in $C(5,15^{\circ})$ collineations 
and the $11A$-TCC clusters at $T=0.5$.
The squares and circles are the results from the simulations and the curves are the Gaussian fits.
The data were obtained from 1000 configurations produced in 10 independent simulation runs.
(b) The PD for the particles  
in the $C(5,15^{\circ})$ collineations 
and the $11A$-TCC clusters simultaneously.
Note that the number of particles in the overlap is rather small: the peak of the PD is located at $\approx 2.8\%$ of 
$N_{A}$.
(c) The PD for the overlap non-randomness parameter 
$\xi_{ov}(t)$. 
Since the  peak of the PD is located at
$\xi_{ov}(t) \approx 0.92$ 
we can say that there is slight avoidance for the particles in the collineations to be also in $11A$-TCC clusters.
}\label{fig:overlap-C515-11A}
\end{center}
\end{figure*}


\section{Discussion of the possible connection with some clusters studied within the Topological Cluster Classification approach \label{sec:Discussion-TCC}}

Relatively recently the structures of 
the KA and Wahnost\"{o}m model 
liquids were studied using 
the topological cluster classification (TCC)
method ~\cite{malins2013identification,malins2013lifetimes,malins2013longlivedclusters}.
The method is based on the idea 
that the geometries of local configurations 
that occur in supercooled liquids
should be closely related to 
the geometries of isolated clusters of particles
with low potential energies.

For the KA liquid, 
one of the low energy configurations 
is the bicapped square antiprism 
or $11A$ clusters. 
It was demonstrated
that these clusters in supercooled liquid
agglomerate into large domains ~\cite{malins2013lifetimes}.
In this context, for example,
see also 
Ref.~\cite{dzugutov2002decoupling,
doye2003favoured,
coslovich2007understanding}.
Thus, there is a possibility that the collineations
might be related to some of such domains.
In order to investigate this possibility,
we studied the non-randomness of the overlap between 
the particles forming the collineations and the particles
forming some clusters from the TCC approach.
Below we describe the adopted method.
Let us suppose that we have two groups of particles. 
In the first group, i.e., the  collineations-group, 
there are particles that form the collineations of a particular type. 
In the second group, i.e., the clusters-group, 
there are particles that form the TCC clusters of a particular type. 
Further, let us assume that the probability, 
for a particular particle that is already in 
the collineations-group
to belong also to the clusters-group does not depend on whether 
this particle belongs to the collineations-group or not. 
Then, 
the probability for an average $A$-particle 
to be 
in the group that represents 
the overlap between the two groups is:
\begin{eqnarray}
P_{ovrl} = P_{coll} \cdot P_{TCC},  \label{eq:chain-tcc-01}
\end{eqnarray}
where
\begin{eqnarray}
P_{coll} \equiv \left \langle \frac{N_{coll}}{N_A} \right \rangle,\;\;\;
P_{TCC} \equiv \left \langle \frac{N_{TCC}}{N_A} \right \rangle
\label{eq:chain-tcc-02}
\end{eqnarray}
are the probabilities for $A$-particle to be in the
collineations and in the clusters groups.
In (\ref{eq:chain-tcc-02}), 
$N_A$ is the total number of 
$A$-particles in the system.

Let us now assume that the overlap is not random and that the particles which are already in one of these two groups have a higher probability to be in another. 
In this case, $P_{ovrl} > P_{coll} \cdot P_{TCC}$.
For example, 
let us assume that $P_{coll} < P_{TCC}$ and that all particles in the collineations, $N_{coll}$, 
also belong to the considered TCC clusters.
In this case, $P_{ovrl} = P_{coll}$. 
On the other hand, if $P_{coll} > P_{TCC}$ then $P_{ovrl} = P_{TCC}$. 
In both cases $P_{ovrl} > P_{coll} \cdot P_{TCC}$.
An opposite extreme example 
is given by the situation when 
particles in one group simply cannot be in another, 
i.e., $P_{ovrl} = 0 < P_{coll} \cdot P_{TCC}$.
In general, if particles in one group avoid being in 
another group then $P_{ovrl} < P_{coll} \cdot P_{TCC}$.
Thus, at every temperature, we can consider the probability distribution of the quantity:
\begin{eqnarray}
\xi_{ov}(t) \equiv \frac{\left( N_{ov}/N_{A} \right)}{\left(N_{coll}/N_{A} \right) \left( N_{TCC}/N_{A} \right)} 
\sim \frac{P_{ov}}{P_{coll}P_{TCC}},
\label{eq:chain-tcc-03}
\end{eqnarray}
where $t$ is the time and 
$N_{ov}$, $N_{coll}$, $N_{TCC}$ change with time.
By definition $\xi_{ov}(t)$, can be larger or smaller than 1.
If the distribution of this quantity (with its average value) 
tends to be larger than 1 then it means that the particles 
in the collineation-group have a tendency to belong also to a TCC - group. 
If this quantity tends to be smaller than 1 then 
the particles avoid being in the collineations and TCC groups 
simultaneously.
%

\begin{figure*}
\begin{center}
\includegraphics[angle=0,width=6.5in]
{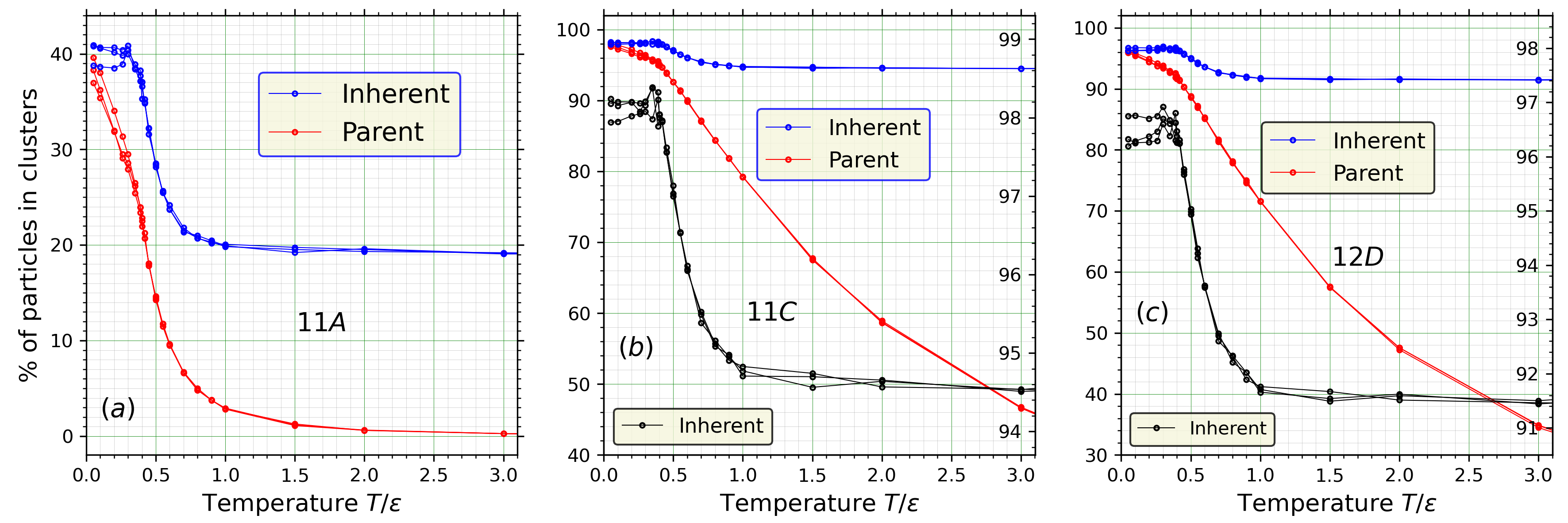}
\caption{
The dependencies of 
the average percentage of both types of particles 
in the clusters of the selected types
on the system's temperature. 
The percentage is with respect to 
the total number of particles 
in the system. 
The results from the inherent structures are also shown.
In panels (b,c) the results from the inherent structures are also shown 
on different scales with the black curves. 
The scales corresponding to the black curves are shown on the right $y$-axes.
The results have been obtained 
using the topological cluster classification analysis
applied to the system containing 8000 particles.
The results from three independent simulations
are shown with the three curves. 
From each simulation 100 configurations 
were used for data analysis. 
}\label{fig:NoP-TCC}
\end{center}
\end{figure*}

%
In Fig.~\ref{fig:overlap-C515-11A}(a) we show the 
probability distributions for the fractions of  
$A$-particles in $11A$-TCC clusters and in $C(5,15^{\circ})$ collineations. 
In Fig.~\ref{fig:overlap-C515-11A}(b) the probability 
distribution for the particles in the overlap is shown.
In Fig.~\ref{fig:overlap-C515-11A}(c) we show the probability distribution 
for the parameter $\xi_{ov}(t)$ (\ref{eq:chain-tcc-03}).
The shown data suggest that there is a slight avoidance for particles in collineations to be also in $11A$-TCC-clusters. 
However, according to panel (b), 
the number of particles in the overlap 
is rather small. Thus, it is not clear 
how much significance can be assigned to this avoidance.

We also studied the overlap between the
$C(5,15^{\circ})$ collineations and several other
TCC clusters. 
The results are summarized in Table~\ref{table:chains-clusters}.
It follows from the data that 
the overlap between the collineations and
the studied TCC-clusters is nearly random.

\begin{table}
\begin{tabular}{| c | c | c | c | c | c | c | c | c | } \hline
Cluster  & $11A$ & $11E$ & $FCC$ & $12B$ & $12D$ & $12E$ & $12K$ & $BCC (9)$  \\ \hline
$P_{coll}$  & $0.21$ & $0.21$ & $0.21$ & $0.21$ & $0.21$ & $0.21$ & $0.21$ & $0.21$  \\ \hline 
$P_{TCC}$  & $0.145$ & $0.99$ & $0.88$ & $0.44$ & $0.91$ & $0.74$ & $0.24$ & $0.82$  \\ \hline  
$\xi_{ov}$  & $0.95$ & $1.01$ & $1.01$ & $1.03$ & $1.05$ & $1.03$ & $0.98$ & $1.00$  \\ \hline 
\end{tabular}
\caption{
Addressing the overlap between 
the $C(5,15^{\circ})$ collineations 
and selected TCC clusters at $T=0.5$.
We see that for all considered clusters 
there is no clear evidence for the presence 
of correlations between the particles that form clusters
and the particles that form collineations.
}
\label{table:chains-clusters}
\end{table}

\subsection{Potential Energy Landscape crossover in the number of particles in selected TCC clusters \label{sec:TCC-crossover}}

In Fig.~\ref{fig:NC-ave-vs-T},\ref{fig:PPDF-and-PADF}(c,d),\ref{fig:Chains-Direct-vs-Model}
it was demonstrated that the numbers of particles 
in collineations and certain parameters
associated with the PDF and ADF exhibit crossovers clearly associated with the crossover in the PEL. 
In this context, it is reasonable to note that these are not the only structural parameters exhibiting this crossover. 
For example, 
in Fig. 2(b) of Ref.~\cite{dzugutov2002decoupling}, 
similar crossover
was observed in the number of icosahedral clusters.
In Fig.~\ref{fig:NoP-TCC}, we show that this crossover can also be observed in the numbers of particles in several TCC clusters.
It follows from the data that 
the PELC is observable for all studied clusters, 
while it is significant only for the $11A$ clusters. 
The case of $11A$ clusters is also
of special interest because for these clusters 
the presence of the crossover is quite obvious 
even from 
the analysis of the parent structures. 
This is in agreement with Ref.~\cite{malins2013lifetimes}
in which 
it has been suggested that 
$11A$ clusters play a special role 
in the dynamic slowdown. 
Thus, the crossover in the number 
of collineations is only one example 
of the development
of structural organization in liquids below the PEL crossover.

\subsection{Diffusion \label{ssec:diffusion}}
%
\begin{figure*}
\begin{center}
\includegraphics[angle=0,width=7.0in]
{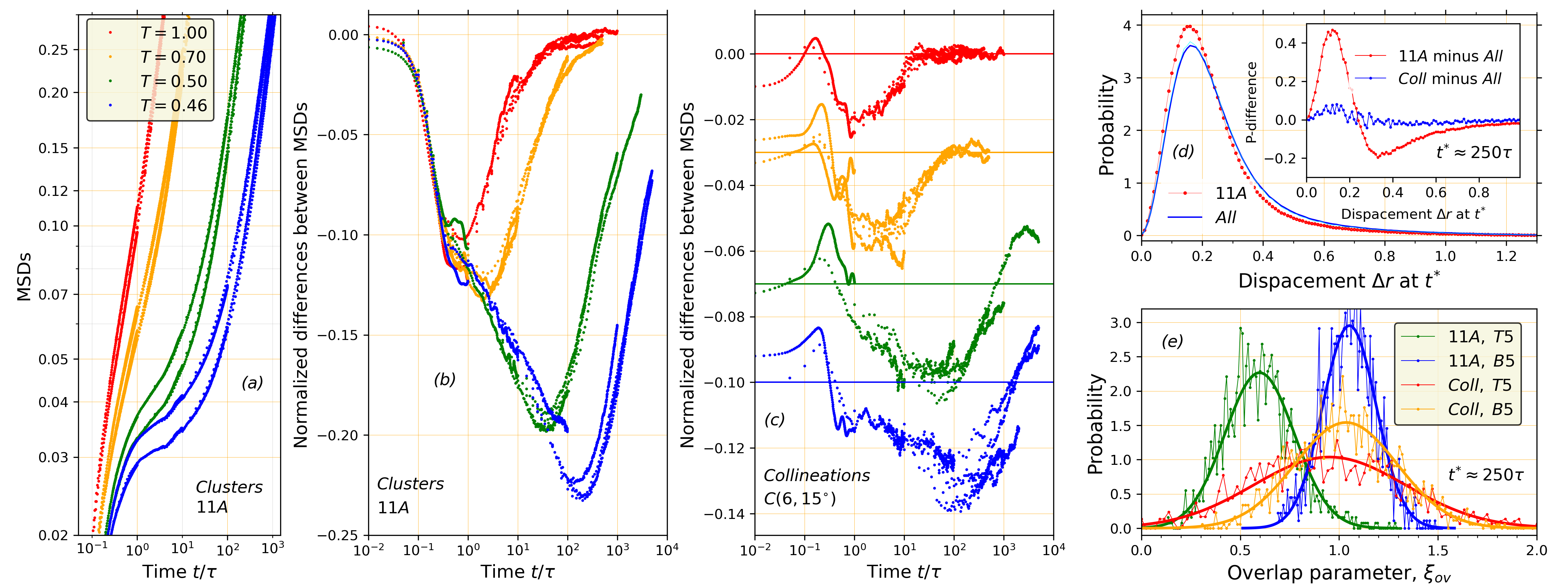}
\caption{
(a) 
The two blue curves 
show 
the dependencies of 
the mean square displacements (MSDs) of
the particles that 
belong 
to the $11A$ clusters 
and those that
do not belong
to the $11A$ clusters
at
$(T/\epsilon) = 0.46$.
The particles that belong to 
the $11A$ clusters diffuse slower, 
i.e., 
to them corresponds the lower blue curve.
The other curves in panel (a) show similar results 
at the other selected temperatures.
If the results for 
the particles that 
belong and do not belong to
the collineations
were plotted in panel (a) 
they would be very close, 
to the results for
the particles that do not belong to the $11A$ 
clusters.
For this reason, 
we do not show these data in panel (a).
(b) The normalized differences 
between the MSDs 
for the particles that belong to the $11A$ clusters 
and the MSDs for the particles that do not 
belong to the $11A$ clusters. 
The normalization is to the MSDs of all particles, 
i.e.,
we plot $\{[MSD(in)-MSD(out)]/MSD(all)\}$.
(c) The results shown in this panel are analogous
to those in panel (b). 
However, these results are for the
$C(6,15^{\circ})$ collineations.
The curves corresponding to the lower
temperatures were displaced downwards.
The horizontal lines of the same color as the data
correspond to zeros for the displaced curves.
Note that the ``dynamic heterogeneity" effect for the $C(6,15^{\circ})$ collineations
is approximately five times smaller than the ``dynamic heterogeneity" 
effect for the $11A$ clusters at all temperatures.
(d) The PDs of the particles' displacements 
(these PDs are also the self-parts of the van Hove correlation function \cite{hansen2013theory}) 
at time $t^{*}$ approximately corresponding to the maximum 
of the non-Gaussian parameter $\alpha_2(t)$, 
i.e., at $t^{*}$ occurs the maximum of the dynamic heterogeneity effect \cite{kob1997dynamical}. 
The red curve in the inset of (d) shows 
the difference between the PDs 
for the $A$-particles that belong to the $11A$ clusters
and the PD for all $A$-particles.
The blue curve shows the difference between the PDs of the particles in $C(6,15^{\circ})$ collineations and the PDs of all particles. 
This difference for the collineations is really small.
In panel (e), the green data show the PD for the overlap parameter $\xi_{ov}$ (\ref{eq:chain-tcc-03}) 
between the top $5\%$ of the mobile particles ($T5$) 
and the $A$-particles in the TCC clusters. 
The maximum of the Gaussian fit curve is close to $0.6$.
This clearly shows that the particles forming the TCC clusters avoid being mobile. 
The blue curve shows the overlap between 
the bottom $5\%$ of the mobile particles 
and the TCC clusters.
The maximum value around $1.1$ shows that
$A$-particles in TCC clusters tend to be slower.
The results for the overlap between the $C(6,15^{\circ})$ collineations and top/bottom $5\%$ of the mobile particles are shown as red/orange curves. 
It is clear that the overlap is essentially random.
}\label{fig:tcc-lll-diffusion}
\end{center}
\end{figure*}

%
One basic question, which is of interest in 
the context of the observed collineations, 
concerns the rate of diffusion of the particles that form the collineations 
and how this rate compares to the rate of diffusion of
other particles. 
In the context of the clusters of particles, for the Wahnstr\"{o}m system,
this question has been addressed 
in Fig.5(c) of Ref.~\cite{malins2013longlivedclusters}.
Here, we follow the same approach for the KA system and compare
the results for the collineations with the results for the $11A$ clusters \cite{malins2013lifetimes}.
In Fig.~\ref{fig:tcc-lll-diffusion} (a)
we show 
the time-dependencies of 
the MSDs
of the particles that belong to the $11A$ and those 
that do not belong to the $11A$ clusters.
As the temperature decreases, the differences between the curves
corresponding to the same temperature increase.
The differences are the largest at the late $\beta$-relaxation time, i.e.,
when the particles, on average, leave their cages.
This behavior can also be observed in panel (b) that shows 
the normalized differences between the curves of the same color in panel (a).
Notably, the difference-curves exhibit well-defined minimums that, 
as the temperature decreases, shift to larger times, as expected.
Note also that, at the lowest temperature, at the minimum, 
the normalized difference is approximately $23\%$ of the MSD calculated
by averaging over all particles.
In panel (c), 
we show the results, 
analogous to those in panel (b), 
but for the
particles that belong and do not belong 
to the $C(6,15^{\circ})$ collineations.
For the clarity of presentation, 
the results for lower temperatures 
(shown in legends of panel (a))
were shifted downwards.
The comparison of the data in (c) with (b) shows that
the slow-down effect for the particles in collineations is approximately
five times smaller than 
for the particles in $11A$ clusters.
The smallness of the effect is the reason 
why we did not plot the MSDs corresponding 
to the collineations in panel (a)--
it would be difficult
to distinguish between 
the collineations-related curves.
Nevertheless, panel (c) clearly shows that the slowdown-effect is there.

In view of the results shown in 
Fig.~\ref{fig:Coordination-Number-CN-Cutoff-and-Voronoi} and 
Fig.~\ref{fig:Common-Neighbor-Analysis-CNA-Cutoff-and-Voronoi},
it can be expected that small differences in the diffusion rates for all particles and the particles in the collineations are caused by small statistical differences in the local environments.

The data shown in Fig.\ref{fig:tcc-lll-diffusion}, for each temperature, 
were accumulated in several simulation runs of different lengths. 
The averaging was done over 1000 initial configurations in each run.

\section{Discussion of the possible connection with the stringlike cooperative motion \label{sec:Discussion-Chain-Like}}

We also investigated if there is a relation between the collineations 
and the stringlike cooperative motion \cite{donati1998stringlike}.

In Ref.~\cite{donati1998stringlike}, 
the attention was focused on $5\%$ 
of the most  mobile particles at time $t^*$, 
which corresponds to the maximum of 
the non-Gaussian parameter 
$\alpha_2(t) = (3<r^4(t)>/5<r^2(t)>^2) - 1$ \cite{kob1997dynamical}.
It was demonstrated that in these $5\%$ 
there are some particles that move in a stringlike fashion
and that, in the studied range of temperatures,
the amount of such particles changes from approximately $50\%$ to $75\%$
out of these $5\%$ of the most mobile particles.

The system that was studied in 
Ref.~\cite{donati1998stringlike,kob1997dynamical}
is the same as the system that we study here.
Thus, to address the relation between the collineations and the
mobile strings, we studied the overlap between the 
particles that form the collineations and the particles 
that form strings. 
The collineations were determined from the static structures at the initial times.
The particles that form strings at the initial time were determined from the consideration of the
particles' displacements in the time window $t \approx t^{*}$ that started at the initial time. 
In total, we considered approximately 1200 initial configurations.

As benchmarks, 
we considered the overlaps between the $5\%$ 
of the most mobile particles and the particles that
formed the $11A$-TCC clusters. 
Besides, 
we considered the overlaps between 
the collineations and TCC clusters with 
the least mobile $5\%$ of the particles.

The results are shown in Fig.~\ref{fig:tcc-lll-diffusion}(d,e).
In Fig.~\ref{fig:tcc-lll-diffusion}(d), 
the cyan curve, corresponding to 
the PD for the displacements of the particles 
in $C(6,15^{\circ})$ collineations, 
nearly coincides with the blue-displacement curve for all particles. The small difference is visible near the peak. 
This difference can be observed also in the inset, in the behavior of the blue curve at $\Delta r \approx 0.1$.
Thus, in Fig.~\ref{fig:tcc-lll-diffusion}(d) we do not observe significant differences in the behaviors of the particles in the collineations from the behavior 
of all $A$-particles.

In Fig.~\ref{fig:tcc-lll-diffusion}(e),
the PDs for several overlap parameters are shown (see Eq.\;\ref{eq:chain-tcc-03}).
It follows from the data that the overlap between the
$11A$-clusters and the mobile particles is clearly non-random, i.e., as expected, the particles in $11A$ clusters avoid being mobile.
Results for the $C(6,15^{\circ})$ collineations show that the overlap between the collineations and the mobile particles is almost random.

Thus, from Fig.~\ref{fig:tcc-lll-diffusion}(d,e) we conclude that
collineations are not related to the mobility strings.

\section{Conclusion \label{sec:conclusion}}

We studied the collineations of more than 
three particles in the KA model system. 
Though the observation of long collineations
was first reported by Bernal  in 1962, 
it appears that since then long collineations
have never been studied systematically.

We found that the average number of particles
in the collineations increases as the temperature of the liquid decreases. 
Below the GT, 
the increase in the number of collineations nearly vanishes.

We also studied collineations 
in the inherent structures.
We found that the average number of particles in the collineations observed on the inherent structures
nearly does not depend on the temperature of the parent liquid 
if it is above the potential energy landscape crossover temperature, $T_x$. 
As the temperature of the parent liquid decrease below $T_x$, 
the number of particles in the collineations observed on the inherent structures starts to increase. 
Thus, the potential energy crossover is reflected also in  the collineations.

Counterintuitively, 
we found that the average number of particles 
in the collineations 
observed on the parent structures 
can be larger 
than the average number of particles in 
the collineations observed on 
the corresponding inherent structures.
It is unclear what is the origin of this situation. 
It is possible that this is associated with the fact 
that the pressure of the inherent structures 
becomes negative when the temperature of 
the corresponding parent liquid 
is lower than a certain value.
In any case, it appears that the relaxation 
from the parent to inherent structures 
destroys some ordering that develops 
in the KA model 
liquid during supercooling.

We also suggested a model that connects 
the numbers of particles 
in the collineations 
with the PDF and ADF.
The model performs rather well.
From this good agreement it could be argued that 
the developments of some features 
in the ADF and PDF are caused by the 
developments in the intermediate 
range order (and beyond).
We also investigated possible connections 
of the collineations with several other phenomena 
studied in the context of supercooled liquids.
In particular, we studied if there are connections with: 
1) the disclinations lines from the geometric frustration approach,
2) selected clusters from the topological cluster classification method of describing the structural changes, 
and 3) the stringlike cooperative motion.
We did not find a connection with any of these phenomena.

Here we did not study a possible connection of 
the collineations with the
nucleation and crystallization processes. 
The KA system is not well-suited for this
investigation. 
Yet, in our view, 
the possibility of such a connection 
deserves separate considerations.

\section{Acknowledgements} 

We are planning to add acknowledgements later.





\clearpage


\title{{\ck Supplemental Materials for the Paper:} \\
{\ck Collineations of particles in the Kob-Andersen system}}

\author{V.A.~Levashov}
\affiliation{Technological Design Institute of Scientific Instrument Engineering, 630055, Novosibirsk, Russia. E-mail: valentin.a.levashov@gmail.com}


\begin{abstract}
In these supplemental materials (SM), 
we describe the interaction potentials 
used in our simulations of the Kob-Andersen system (KAS) 
and provide the details of our simulation procedure.
We also describe the algorithm used for the chain-search. 
Besides, we show the data on the pressure of inherent 
structures and how it depends on 
the temperature of the parent structures.
Finally, these SM contain the figures analogous to Fig.5,6
in the main text, but calculated on the parent structures.
\end{abstract}



\today
\maketitle

\tableofcontents


\section{The used potential}

\begin{center}
\begin{table*}
\begin{tabular}{| c | c | c | c | c |} \hline
$T$             & $dt$           & $\Delta t$  & $<[\Delta r_A(\Delta t)]^2>,\;<[\Delta r_B(\Delta t)]^2>$   & $\tau_{\alpha}$ \\ \hline
$6.0$           & $0.0005 \tau$  & $10\tau$    & $\approx 12.1\sigma^2$, $\approx 17.3\sigma^2$              &   no data   \\ \hline
$5.0$           & $0.0010 \tau$  & $20\tau$    & $\approx 20.4\sigma^2$, $\approx 28.1\sigma^2$              &   no data   \\ \hline
$4.0$           & $0.0010 \tau$  & $10\tau$    & $\approx 7.7\sigma^2$, $\approx 11.0\sigma^2$               &   no data   \\ \hline
$3.0$           & $0.0010 \tau$  & $20\tau$    & $\approx 15.6\sigma^2$, $\approx 21.5\sigma^2$              &  $\approx 0.23 \tau$   \\ \hline
$2.5$           & $0.0010 \tau$  & $20\tau$    & $\approx 8.2\sigma^2$, $\approx 12.1\sigma^2$               &  $\approx 0.28 \tau$   \\ \hline
$2.0$           & $0.0010 \tau$  & $20\tau$    & $\approx 5.7\sigma^2$, $\approx 8.5\sigma^2$                &  $\approx 0.36 \tau$   \\ \hline
$1.6$           & $0.0010 \tau$  & $20\tau$    & $\approx 3.9\sigma^2$, $\approx 6.2\sigma^2$                &  $\approx 0.49 \tau$   \\ \hline
$1.5$           & $0.0010 \tau$  & $30\tau$    & $\approx 5.2\sigma^2$, $\approx 8.1\sigma^2$                &   no data   \\ \hline
$1.4$           & $0.0010 \tau$  & $50\tau$    & $\approx 7.6\sigma^2$, $\approx 11.1\sigma^2$               &   no data   \\ \hline
$1.2$           & $0.0010 \tau$  & $50\tau$    & $\approx 5.5\sigma^2$, $\approx 8.3\sigma^2$                &  $\approx 0.80 \tau$   \\ \hline
$1.0$           & $0.0020 \tau$  & $100\tau$   & $\approx 6.8\sigma^2$, $\approx 10.8\sigma^2$               &  $\approx 1.19 \tau$   \\ \hline
$0.8$           & $0.0020 \tau$  & $100\tau$   & $\approx 3.4\sigma^2$, $\approx 5.8\sigma^2$               &   $\approx 2.39 \tau$   \\ \hline
$0.7$           & $0.0030 \tau$  & $150\tau$   & $\approx 3.0\sigma^2$, $\approx 5.2\sigma^2$               &   $\approx 4.49 \tau$   \\ \hline
$0.6$           & $0.0030 \tau$  & $150\tau$   & $\approx 3.0\sigma^2$, $\approx 5.2\sigma^2$               &   $\approx 12.7 \tau$   \\ \hline
$0.55$          & $0.0030 \tau$  & $300\tau$   & $\approx 1.25\sigma^2$, $\approx 2.43\sigma^2$              &  $\approx 28.3 \tau$   \\ \hline
$0.50$          & $0.0030 \tau$  & $300\tau$   & $\approx 0.37\sigma^2$, $\approx 0.94\sigma^2$              &  $\approx 100.7 \tau$   \\ \hline
$0.48$          & $0.0050 \tau$  & $500\tau$   & no data                                                     &  $\approx 222.0 \tau$   \\ \hline
$0.47$          & $0.0050 \tau$  & $500\tau$   & no data                                                     &  $\approx 360.8 \tau$   \\ \hline
$0.46$          & $0.0050 \tau$  & $500\tau$   & $\approx 0.22\sigma^2$, $\approx 0.63\sigma^2$              &  $\approx 642 \tau$   \\ \hline
$0.45$          & $0.0050 \tau$  & $500\tau$   & $\approx 0.137\sigma^2$, $\approx 0.41\sigma^2$              &  $\approx 1'265 \tau$   \\ \hline
$0.44$          & $0.0050 \tau$  & $500\tau$   & $\approx 0.122\sigma^2$, $\approx 0.35\sigma^2$              &  $\approx 2'313 \tau$   \\ \hline
$0.43$          & $0.0050 \tau$  & $500\tau$   & $\approx 0.078\sigma^2$, $\approx 0.24\sigma^2$              &  $\approx 4'945 \tau$   \\ \hline
$0.42$          & $0.0050 \tau$  & $500\tau$   & $\approx 0.0514\sigma^2$, $\approx 0.15\sigma^2$              &  $\approx 10'065 \tau$   \\ \hline
$0.41$          & $0.0050 \tau$  & $500\tau$   & $\approx 0.044\sigma^2$, $\approx 0.112\sigma^2$              &   no data   \\ \hline
$0.40$          & $0.0050 \tau$  & $500\tau$   & $\approx 0.035\sigma^2$, $\approx 0.094\sigma^2$              &   no data   \\ \hline
$0.39$          & $0.0050 \tau$  & $500\tau$   & $\approx 0.029\sigma^2$, $\approx 0.068\sigma^2$              &   no data   \\ \hline
$0.38$          & $0.0050 \tau$  & $500\tau$   & $\approx 0.028\sigma^2$, $\approx 0.071\sigma^2$              &   no data   \\ \hline
$0.37$          & $0.0050 \tau$  & $500\tau$   & $\approx 0.025\sigma^2$, $\approx 0.045\sigma^2$              &   no data   \\ \hline
\end{tabular}
\caption{
The purpose of 
the table 
is to provide 
some technical details  about
the simulation procedure. 
Thus, 
the 1st column shows 
some temperatures at which 
the simulations have been performed. 
The 2nd column 
shows 
the values of 
the time step 
used at 
the corresponding temperatures. 
We generated 
1000 configurations of 
8000 particles in
10 independent runs 
at 
each studied temperature. 
Thus, 
in each independent run, 
we generated 100 configurations. 
The time interval 
between 
consecutive structures 
saved for later analysis 
is provided in the 3rd column.
In the 4th column, 
we show 
the mean square particles' displacements
corresponding to 
the time between 
two consecutively saved structures.
Note that 
for temperatures 
$T \geq 0.55 \epsilon$ 
the time interval 
between 
two consecutively saved structures 
is such that 
the mean square displacement of 
$A$-particles 
during this time interval 
is larger than $\sigma^2$. 
Thus, 
it is natural to consider 
such structures as 
quite different.
For such temperatures, 
this time interval is 
also significantly larger 
than 
the $\alpha$-relaxation time.
For the temperatures in 
the interval 
$0.46 \le T \leq 0.47$ 
the dynamics is 
already quite slow and 
at these temperatures 
the MSDs 
corresponding to the time intervals 
between consecutively saved structures 
are smaller than $\sigma$.
However, 
the time intervals 
between 
the two consecutively saved structures 
are still larger than 
than the corresponding
$\alpha$-relaxation times
that we show in the 5th column.
For temperatures 
$T \leq 0.46$ 
the $\alpha$-relaxation time 
becomes 
quite large.  
For such low temperatures 
we did not aim 
to produce 
truly independent configurations.
Instead, 
our purpose was 
to investigate how 
further structural relaxation 
affects the quantity of chains, 
even thought 
it is clear that 
we do not study 
the true equilibrium and 
truly independent 
configurations. 
Actually, 
precisely for this reason, i.e.,
to have more independent configurations, 
we, 
starting from 
the very high temperatures, 
utilized 10 independent simulation runs.
In any case, 
for $T < 0.46$ 
we adopted 
the following relaxation procedure. 
After reducing 
the temperature of 
the system from 
the previous temperature above, 
we relaxed the system 
for 
$t = 50'000\tau$.
After this relaxation, 
we  collected 
100 configurations 
during the same time 
$t = 50'000\tau$.
The final configuration of 
the particles, 
after the data collection run, 
was used as 
the starting configuration for 
the next lower temperature. 
This 
was done for 
each of the 10 independent runs. 
In this way, 
we reduced 
the temperature 
down to $T=0.05\epsilon$.
For the larger system of 
64'000 particles, 
we followed the same procedure for 
those temperatures that we studied.
}
\label{table:phase-regions}
\end{table*}
\end{center}

\begin{figure*}
\begin{center}
\includegraphics[angle=0,width=6.6in]{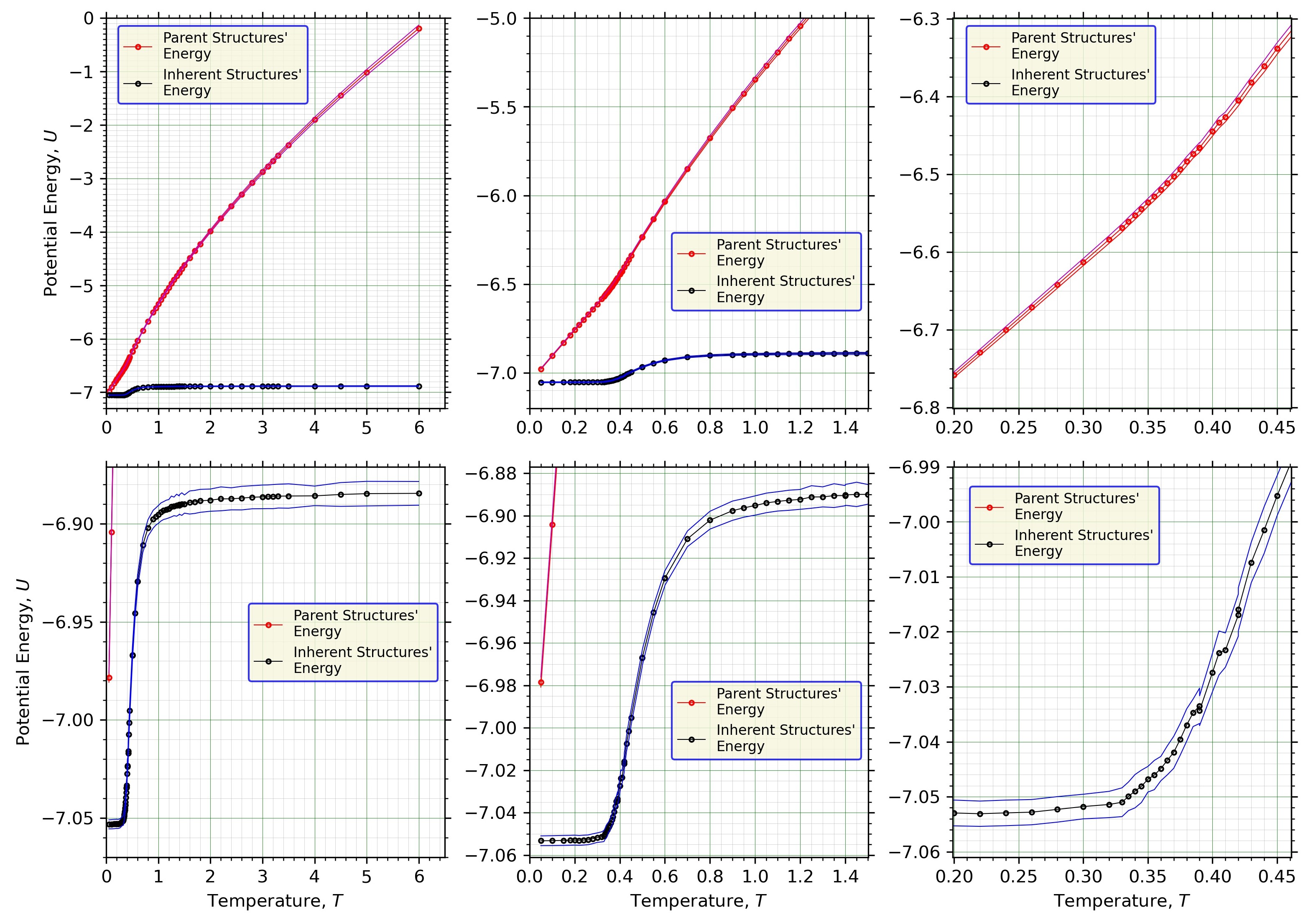}
\caption{
Different panels 
show 
the same data on 
different scales. 
In all panels, 
the red curves 
show how 
the total average potential energy of 
the system 
per particle 
depends on the temperature. 
In every panel 
there are 3 red curves--the average 
and
the average $\pm \sigma$, 
where 
$\sigma=\sigma(T)$ 
is the width of 
the distribution of 
the potential energies of 
individual structures
(it is not the $\sigma$ of the mean).
The black curves in 
all panels 
show 
the temperature dependence of 
the average inherent structures' potential energy.
The blue curves 
correspond to 
the black curves $\pm \sigma$ of 
the distribution of 
the inherent energy values of 
individual structures 
(it is not the $\sigma$ of the mean).
}\label{fig:PE-parent-inherent}
\end{center}
\end{figure*}

\begin{figure*}
\begin{center}
\includegraphics[angle=0,width=6.6in]{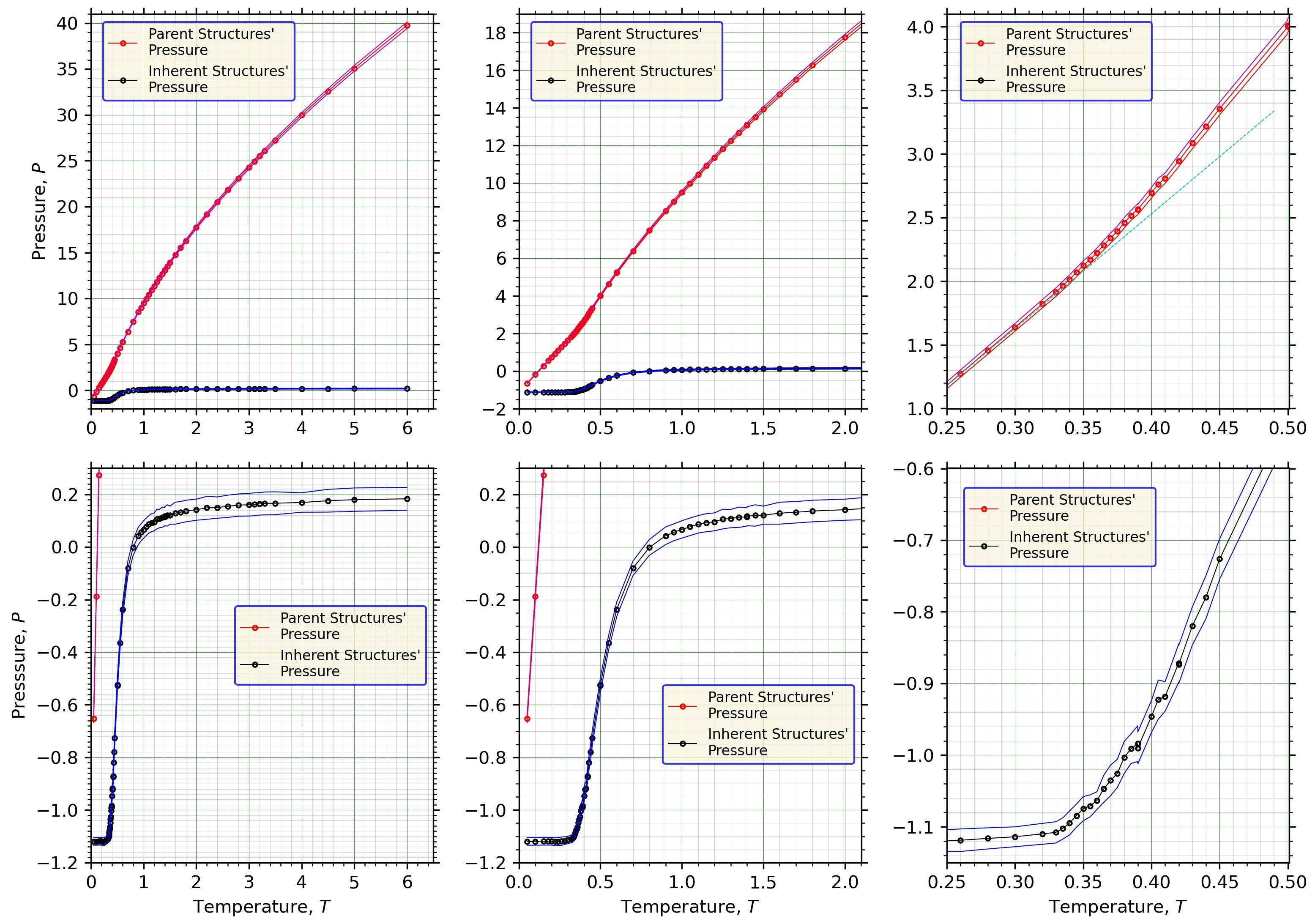}
\caption{
Different panels 
show 
the same data on 
different scales. 
In all panels, 
the red curves 
show how 
the average pressure of 
the parent systems 
depends on the temperature. 
In fact, 
in every panel there are 
3 red curves--the average 
and the average $\pm \sigma$, 
where $\sigma=\sigma(T)$ is 
the width of the pressure distribution of 
individual structures
(it is not the $\sigma$ of the mean).
The black curves in 
all panels 
show 
the temperature dependence of 
the inherent structures' average pressures.
The blue curves 
correspond to 
the black curves 
$\pm \sigma$ of 
the distribution 
of the energy values from 
individual structures 
(it is not the $\sigma$ of the mean).
}\label{fig:PP-parent-inherent}
\end{center}
\end{figure*}

%
In our simulations, 
we used 
the Kob-Andersen 
binary model system of particles 
\cite{1995KobAndersen01,1995KobAndersen02}.
This system consists of 
$80\%$ 
of larger $A$ particles 
and $20\%$ 
of smaller $B$ particles. 
The masses, $m$, of 
all particles are the same.
The particles interact via 
the modified Lennard-Jones potentials 
\cite{stoddard1973numerical,1998SastryPEL,2002KApressureBagchi,2013motifsMalinsTanaka}. 
The modifications guarantee that 
the interactions between 
the particles of types $a$ and $b$ 
become zero at $2.5\sigma_{ab}$ 
\cite{stoddard1973numerical}. 
The continuities of 
the derivatives of the potentials (the forces) 
at $2.5\sigma_{ab}$ are also guaranteed.
The precise forms of 
the potentials 
are given by 
\cite{stoddard1973numerical,2002KApressureBagchi}:
\begin{eqnarray}
U_{ab}(r) = 4 \epsilon_{ab} \left\{ \phi_{ab}^{LJ}(r) + \phi_{ab}^{D}(r) + C_{ab} \right\},
\label{eq:potential-01}
\end{eqnarray}
where
\begin{eqnarray}
&&\phi_{ab}^{LJ}(r) = \left(\frac{\sigma_{ab}}{r}\right)^{12} - \left(\frac{\sigma_{ab}}{r}\right)^{6},\label{eq:potential-02-1}\\
&&\phi_{ab}^{D}(r) = \left[6\left(\frac{\sigma_{ab}}{r_{abc}}\right)^{12}
                 - 3\left(\frac{\sigma_{ab}}{r_{abc}}\right)^{6}\right]\left(\frac{r}{r_{abc}}\right)^2,\label{eq:potential-02-2}\\
&&C_{ab} = \left[7\left(\frac{\sigma_{ab}}{r_{abc}}\right)^{12}
                 - 4\left(\frac{\sigma_{ab}}{r_{abc}}\right)^{6}\right],\label{eq:potential-02-3}
\end{eqnarray}
where  
$\phi_{ab}^{LJ}(r)$ 
(\ref{eq:potential-02-1}) 
are the Lennard-Jones parts 
of the potentials. 
The terms 
$\phi_{ab}^{D}(r)$ 
(\ref{eq:potential-02-2}) 
guarantee 
the continuities
of the derivatives 
of the potentials at 
the cutoff distance $r=r_{abc}=2.5\sigma_{ab}$ 
(the subscript ``$c$" in $r_{abc}$ stands for ``cutoff").
The constants 
$C_{ab}$ 
(\ref{eq:potential-02-3}) 
guarantee 
the continuities of 
the potentials at $r=r_{abc}$.

The length scales of 
the interactions 
between the particles 
of different types are: 
$\sigma_{AA} = \sigma = 1.0 \sigma $, 
$\sigma_{BB} = 0.88 \sigma$, and 
$\sigma_{AB} = 0.80 \sigma$. 
Note that $\sigma_{AB} \neq (\sigma_{AA} + \sigma_{BB})/2 = 0.90$. 
Though, the difference between $0.88$ and $0.90$ is quite small.
The energy scales are: 
$\epsilon_{AA} = \epsilon = 1.0 \epsilon$, 
$\epsilon_{BB} = 0.5 \epsilon$, and 
$\epsilon_{AB} = 1.5\epsilon$
The units of length and 
energy/temperature 
are $\sigma$ and $\epsilon$. 
The unit of pressure is 
$\epsilon/\sigma^3$.
The unit of time is 
$\tau = \sqrt{m \sigma^2/\epsilon}$.
We note that in 
Ref.~\cite{1995KobAndersen01,1995KobAndersen02} 
a rescaled (a smaller) 
unit of time 
has been used 
$\tau_{KA} = (1/\sqrt{48})\tau$.
On the other hand, 
the unit of time used, 
for example, 
in 
Ref.\,\cite{1998SastryPEL,2002KApressureBagchi,2013motifsMalinsTanaka} 
is the same as 
the one used in our work.

\section{Details of the simulation procedure}

In our simulations, 
we used 
the NVT ensemble 
within the LAMMPS program 
\cite{Plimpton1995,thompson2022lammps,lammps}.
The very initial structure, 
from which we started all simulations,  
was the simple cubic lattice 
with the average density 
of all particles $\rho_o = 1.2$. 
In this lattice, 
4 planes of $A$ particles 
alternated with 
1 plane of $B$ particles. 
Then, we ``melted" this 
artificially created lattice 
within LAMMPS at temperature $T=7.0$. 
At this high temperature, 
the particles are mobile and 
it does not take much time 
to achieve 
the mean square displacement of 
particles 
corresponding to dozens of interatomic distances.
We performed several of such 
initial consecutive equilibration runs.
Then, still at $T=7.0$, we performed several consecutive runs in which the final structures for the restart files were separated from each other by time intervals corresponding to the MSDs corresponding to several dozens of interatomic distances.
Thus, 
we assume that 
we produced several 
sufficiently different structures 
at $T=7.0$. 
For the system of 
8000 particles, 
we produced 10 
structurally different 
restart files for further use. 
For the system of 64000 particles, 
we produced 6 
independent restart files.

Then, 
concerning the systems of 8000 particles, 
for each of the created 
10 configurations, 
we abruptly decreased 
the temperature down to $T=6.0$.
Then, 
each of these 10 configurations 
was equilibrated, 
at $T=6.0$, 
for the time 
corresponding to 
the mean average displacement of particles equal to 
approximately 25 interatomic distances, i.e., 
($t \approx 500\tau$), 
where $\tau = \sqrt{m\sigma^2/\epsilon}$. 
The used value of the time step was 
$dt = 0.0005 \tau$.
Then, 
starting from 10 structures 
equilibrated  at $T=6.0$, 
we generated 100 configurations 
for each 
of the initially 
equilibrated structures. 
Below, 
we address the time intervals 
between the saved structures 
in details. 
Thus, in total, 
we had 1000 configurations 
for the analysis.
Further, 
the procedure described above 
for $T=6.0$ 
was used to obtain the structures 
at lower temperatures. 
The transition to 
a lower temperature 
was performed by 
reducing the temperature from 
the available 
previous higher temperature 
just above.

This procedure 
allows to achieve 
a further configurational divergence of 
the structures obtained from 
the 10 initial independent structures 
at $T=7.0$ 
(in practice, 
it does not matter 
because 
these initial 10 structures at 
$T=7.0$ 
are already very different).
For the temperatures $T > 1.0$, 
the diffusion process is fast -- 
the particles leave their cages 
within times $t < 0.5 \tau$.
Thus, for $T > 1.0$, 
it is not difficult 
to produce the equilibrated systems 
and 
then to produce 
the particles' configurations
``separated" by 
mean particle displacement of 
several $\sigma$, 
i.e., 
presumably, 
quite different configurations.
Note, 
the displacement of 
several $\sigma$ 
guarantees that 
the ``separation" time 
is larger than 
the $\alpha$-relaxation time.
In the temperature interval 
between 
$T=5.0$ and $T=1.0$ 
the used value of the time step 
was 
$dt = 0.001\tau$. 
In 
the temperature interval 
between 
$T=1.0$ and $T=0.5$ 
we increased 
the time step 
from $dt=0.001 \tau$ 
to $dt =0.005 \tau$.
For 
$T \leq 0.5$ 
the time step 
was $dt = 0.005$.
Further information, 
concerning the details of 
the simulation procedure, 
we provide in 
Table \ref{table:phase-regions}.
An insight into the degree of the system
relaxation can also be gained from  
Fig.\,\ref{fig:PE-parent-inherent},\ref{fig:PP-parent-inherent}.

\section{Algorithm for the chain-search}\label{sec:chain-search}

To find 
the  collineations of $A$-particles, 
we used 
the following algorithm:\\
1) In the most external cycle 
(zero-order cycle) 
over all particles in 
the system 
we chose 
a particle ``A" and 
label it as $A_o$.\\
2) We determine 
nearest neighbors of $A_o$ 
according to 
the cutoff distance 
(the 1st minimum in the $AA$-partial PDF).
It is convenient, 
in a preliminary analysis, 
to determine 
the lists of neighbors of 
all $A$-particles.\\ 
3) We run 
the 1st order internal cycle 
over the neighbors of $A_o$.\\
4) A particular chosen $A$-neighbor of $A_o$ 
we label as $A_1$.\\
5) For the chosen 
$A_o$ and $A_1$ 
we run again 
the cycle over 
the neighbors of 
$A_o$ 
(the 2nd order internal cycle).
Another (different from $A_1$) particular 
chosen $A$-neighbor of $A_o$ 
we label as $A_2$.\\
6) For the selected 
$A_o$, $A_1$, and $A_2$, 
we check the value of 
the angle 
$\angle(A_1 A_o A_2)$. 
If 
this angle is 
sufficiently close to 
$180^{\circ}$, 
according to 
a chosen angular cutoff, 
we say that 
there is a chain of 
the 3rd order.\\
7) If 
we are interested in 
chains of the 4th order, 
then 
the particle $A_1$ is used 
as the particle $A_o$ previously.
In particular, 
then 
there is 
a cycle over 
the 1st neighbors of $A_1$ 
(we label them as $A_3$) and 
we check 
the angles $\angle(A_o A_1 A_3)$.\\
8) If 
we are interested in 
the chains of 5th order, 
then, 
after the step 7) of 
this algorithm is completed, 
the particle $A_2$ is 
used as 
the particle $A_o$ previously.
In particular, 
then 
there is a cycle 
over the 1st neighbors of $A_2$ 
(label them as $A_4$) 
and the angle 
$\angle(A_o A_2 A_4)$ 
is checked for 
the angle-cutoff condition.\\
9) If 
we are interested in 
the chains of higher order 
then we introduce 
more cycles 
over the neighbors of 
the particles 
at the ends of 
already determined chains 
of lower order.
%

\begin{figure*}
\begin{center}
\includegraphics[angle=0,width=6.6in]{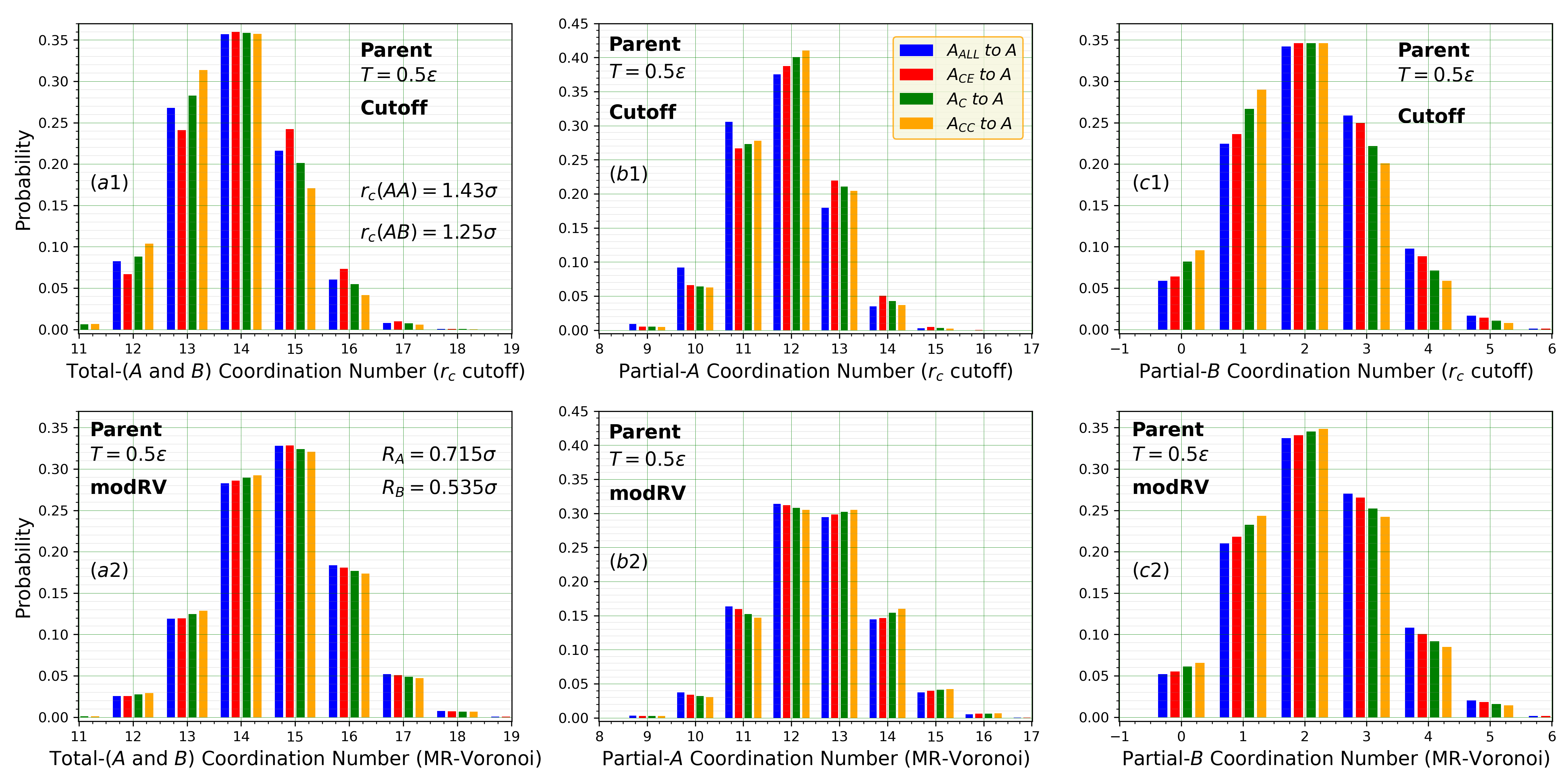}
\caption{
Nearest neighbor analysis of $A$-particles in the collineations.
This figure is completely analogous to Fig.\,5 of the paper.
However, the data in it have been produced from the parent structures at $T=0.5\epsilon$.
}\label{fig:nearest-neighbors-parent}
\end{center}
\end{figure*}
\begin{figure*}
\begin{center}
\includegraphics[angle=0,width=6.6in]{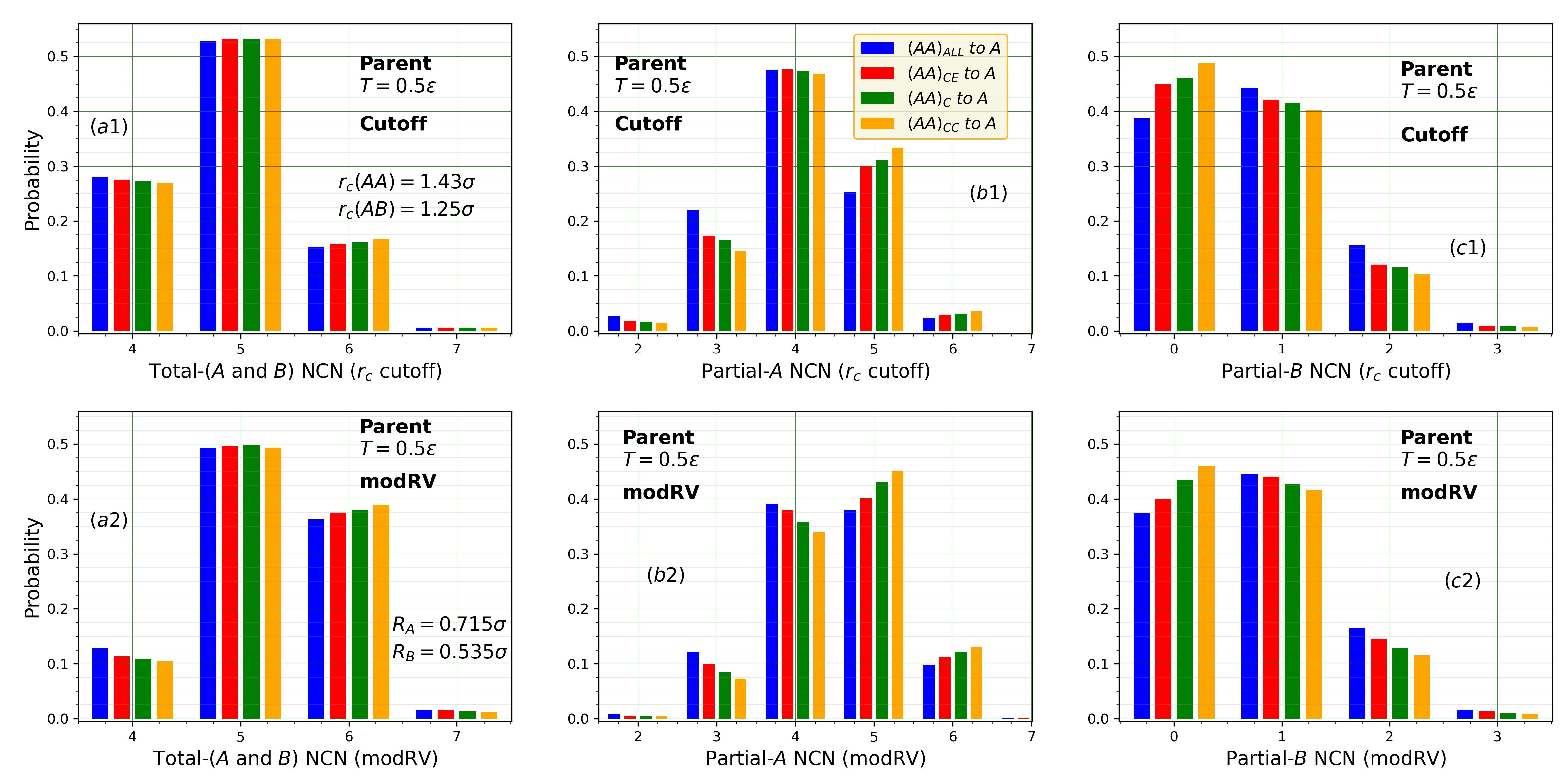}
\caption{
Common neighbor analysis of $AA$-bonds in the collineations.
This figure is completely analogous to Fig.\,6 in the paper.
However, the data in it have been produced from the parent structures at $T=0.5\epsilon$.
}\label{fig:common-neighbors-parent}
\end{center}
\end{figure*}

\section{Potential energies and pressures of the parent and inherent structures}

In Fig.~\ref{fig:PE-parent-inherent},
we show 
how 
the potential energies 
per particle of 
the parent and inherent structures 
depend on the temperature.
In all panels, 
the same data are presented, 
but on different scales.
The results 
shown in this figure 
can be compared, 
for example, 
with the data in
Fig.~1 of Ref.\cite{1998SastryPEL}.
To produce 
the inherent structures we used the FIRE 
algorithm \cite{bitzek2006structural,guenole2020assessment} 
within the LAMMPS program \cite{Plimpton1995,lammps}.
The criterion 
for convergence of 
the minimization process 
was: 
the change in 
the total potential energy in 
a particular step has to be smaller
than $10^{-4}$ of 
the energy value. 
A maximum number of 
$10^4$ steps was allowed.
The value of 
the time step used was 
$dt = 0.002$.
The recommended time step for 
the FIRE algorithm is the same 
as the time step used in 
MD simulations
\cite{lammps}.
Every point in 
the shown plots 
originates from 
1000 configurations of 
8000 particles.
These 
1000 configurations originate from 
10 independent runs, 
as described in 
this SM in 
the section on 
simulation procedure.
In Fig.~\ref{fig:PE-parent-inherent}, 
as 
in Fig.~1 of 
Ref.\cite{1998SastryPEL}, 
we see that
for $T>1.5$ 
the average energies of 
the inherent structures, 
essentially, 
do not depend on 
the temperature.
Then, 
as the temperature of 
the parent system decreases, 
at $T \approx 1.0$, 
there occurs 
a crossover to 
the potential energy landscape (PEL) 
influenced regime.
Then, 
the figure shows that, 
at $T \approx 0.36$, 
the timescale of 
our relaxation procedure of 
the parent structures 
is not sufficient 
to capture the structural 
relaxation of the system.
In Fig.\,\ref{fig:PP-parent-inherent}, 
we show the temperature dependencies of the pressure of parent and inherent structures. 
In principle, 
the behavior of all shown curves 
is similar to the behavior of 
the curves shown in Fig.\,\ref{fig:PE-parent-inherent}. 
Note, that the pressure
of the inherent structures becomes negative at $T \approx 0.8 \epsilon$.
Thus, 
at $T<0.8\epsilon$, 
the inherent structures 
may 
exhibit a tendency for 
the formation of cavities.
The issue about the cavities and 
inhomogeneities in the inherent structures
has been discussed in the past.
See, for example, 
Ref.\,\cite{sastry1997statistical,sastry2000liquid,hernandez2003density,
altabet2016cavitation,altabet2018cavitation,makeev2018distributions}.
In particular, for the Kob-Andersen system, it has been shown that, 
for the reduced densities $\rho \sigma^3 > 1.08$, the inherent structures 
remain homogeneous even when the pressure 
in these structures is negative (tensile).
Thus, the crossover in the number of chains that we observe 
on the inherent structures may not be associated 
with the formation of cavities.

\section{Nearest neighbor and common neighbor analyses of the parent structures}

In Fig.\,\ref{fig:nearest-neighbors-parent},
we show the results of the nearest neighbor analysis conducted on the parent structures
at $T=0.5\epsilon$. 
These results are completely analogous to the results presented in Fig.\,5
of the paper.

In Fig.\,\ref{fig:common-neighbors-parent},
we show the results of the common neighbor analysis conducted on the parent structures
at $T=0.5\epsilon$. 
These results are completely analogous to the results presented in Fig.\,6
of the paper.


\bibliographystyle{zbib-unsrtnat}

\providecommand{\noopsort}[1]{}\providecommand{\singleletter}[1]{#1}%


\end{document}